\documentclass[universe,article,accept,moreauthors,pdftex]{Definitions/mdpi} 

\firstpage{1} 
\pubvolume{6}
\issuenum{3}
\articlenumber{46}
\pubyear{2020}
\copyrightyear{2020}
\datereceived{7 February 2020} 
\daterevised{9 March 2020} 
\dateaccepted{9 March 2020} 
\datepublished{18 March 2020} 
\hreflink{https://doi.org/10.3390/universe6030046} 

\usepackage{booktabs} 
\usepackage{multirow}
\usepackage{soul} 
\usepackage{microtype}

\setitemize{parsep=6pt,itemsep=0pt,leftmargin=*,labelsep=5.5mm,align=parleft}  
 \setenumerate{parsep=6pt,itemsep=0pt,leftmargin=*,labelsep=5.5mm,align=parleft}
\setlist[description]{itemsep=0mm}   

\Title{The Scale-Invariant Vacuum (SIV) Theory: A Possible Origin of Dark Matter and Dark Energy}
\Author{Andre Maeder\orcidA{}  $^{1}$* and Vesselin G. Gueorguiev\orcidB{} $^{2,3}$}

\AuthorNames{Andre Maeder and Vesselin G. Gueorguiev}

\address{%
$^{1}$ \quad Geneva Observatory, University of Geneva, chemin des Maillettes 51, 
CH-1290 Sauverny, Switzerland; Andre.Maeder@unige.ch\\
$^{2}$ \quad Institute for Advanced Physical Studies, Montevideo Street, Sofia 1618, Bulgaria;
Vesselin@MailAPS.org\\ 
$^{3}$ \quad Ronin Institute for Independent Scholarship, 127 Haddon Pl., Montclair, NJ 07043, USA 
}

\corres{{Correspondence}:andre.maeder@unige.ch }

\abstract{The Scale Invariant Vacuum (SIV)
theory rests on the basic hypothesis that the macroscopic empty space is scale invariant.
This hypothesis is applied in the context of the Integrable Weyl Geometry, 
where it leads to considerable simplifications in the scale covariant cosmological equations.
After an initial explosion and a phase of braking, 
the cosmological models show a continuous acceleration of the expansion.
Several observational tests of the SIV cosmology are performed: 
on the relation between $H_0$ and the age of the Universe, 
on the $m-z$ diagram for SNIa data and its extension to $z=7$ with quasars and GRBs, and on the $H(z)$ vs. $z$ relation. 
All comparisons show a very good agreement between SIV predictions and observations.
Predictions for the future observations of the redshift drifts are also given.
In the weak field approximation, the equation of motion contains, 
in addition to the classical Newtonian term, an acceleration term (usually very small) depending on the velocity. 
The two-body problem is studied, showing a slow expansion of the classical conics. 
The new equation has been applied to clusters of galaxies, to rotating galaxies 
(some proximities with Modifies Newtonian Dynamics, MOND, are noticed), 
to the velocity dispersion vs. the age of the stars in the Milky Way, 
and to the growth of the density fluctuations in the Universe.
We point out the similarity of the mechanical effects of the SIV hypothesis in cosmology 
and in the Newtonian approximation. 
In~both cases, it results in an additional acceleration in the direction of motions. 
In cosmology, these effects are currently interpreted in terms of the dark energy hypothesis, 
while in the Newtonian approximation they are accounted for in terms of the dark matter (DM) hypothesis. 
These hypotheses appear no longer necessary in the {SIV context.}
}

\begin{document}

\tableofcontents 



\section{Introduction}

This work pertains to the exploration of new ways to try to understand the dark components of the Universe. 
We present the basic physical assumptions of the Scale-Invariant Vacuum (SIV) theory, 
the cosmological properties with new results, and several tests in the weak field approximation.~The results differ from those of General Relativity (GR) by a small additional acceleration in the sense of motion.
The cosmological solutions are close to the Concordance Model ($\Lambda$CDM), 
but with some moderate time-dependent differences. 

Among the numerous theoretical attempts to explain the dark energy and dark matter in the Universe, the scale-invariant vacuum (SIV)
theory is likely the one based on the most ``economic'' hypothesis, i.e., the assumption that the macroscopic empty space 
is invariant to scale transformations. 
We may call it the most economic one because this hypothesis is first 
a simple minimal extension of Einstein's GR,
compared to the many ones proposed to account for the dark components of the Universe, and, 
as further discussed below, this hypothesis is already satisfied in electrodynamics, 
as well in General Relativity in the absence of cosmological constant.
Over the last few decades, an
impressive variety of new models based on exotic particles and fields have been forwarded in
order to account for the assumed dark components of our Universe. 
Explorations in high energy particle physics have been pursued without much advance in solving the problem;
see recent reviews by~\citet{Bertone17} and\mbox{~\citet{Swart17}}. 
Simultaneously, theories of modified gravity have been developed, 
in particular the theory of Modified Newtonian Dynamics (MOND) developed by Milgrom
\citep{Milgrom83,Milgrom09,Milgrom19}, which 
well accounts for some properties of galactic dynamics; see 
\citet{Famaey12},~\citet{Kroupa12}, \citet{Kroupa15} and~\citet{McGaugh16}.

As pointed out by~\citet{Dirac73}, ``It appears as one of the fundamental principles in Nature that
the equations expressing basic laws should be invariant under the
widest possible group of transformations''. For a long time, there has been a particular concern about a possible scale or gauge 
invariance in the theory of gravitation; see, for example,~\citet{Weyl23},~\citet{Dirac73} and~\citet{Canu77}.

Scale invariance means that the basic equations do not change upon a transformation of the line element of the form, 
\begin{equation}
ds'\,=\,\lambda(t)\,ds\, ,
\label{ds}
\end{equation}
where $\lambda(t)$ is the scale factor, considered here not to depend on space
for reason of homogeneity and isotropy.
We know that a general scale invariance of the physical laws is prevented 
by the presence of matter, which defines scales of mass, time, and length~\citep{Feynman63}. 
However, empty space at large scales could have the property of scale invariance, 
since by definition there is nothing to define a scale in~it. 
As a matter of fact, the scale invariance of the physical laws in empty space is a fundamental and well known property.
Maxwell's equations are scale invariant in empty space. Although it is not so frequently mentioned,
it is also true for General Relativity (GR) if the Einstein cosmological constant 
$\Lambda_{\mathrm{E}}$ is equal to zero. However, the equations of GR in empty space are no longer 
scale invariant if the cosmological constant is different from zero. 
According to~\citet{Bondi90}, this was one of the reasons for Einstein's disenchantment 
with $\Lambda_{\mathrm{E}}$.

The inclusion of scale invariance, in addition to the general covariance, 
implies to move from the Riemann Geometry to Weyl's Geometry, 
which accounts for the possibility of gauge transformations. 
Weyl's Geometry is based on the
co-tensor calculus rather than tensor calculus, since the mathematical objects
such as derivatives of scalars, vectors and tensors, Christoffel's symbols, etc.
must have the property of gauge invariance in addition to the usual general covariance.
Moreover, we use here a particular form of Weyl's Geometry, 
the Integrable Weyl's Geometry, see Appendix~\ref{hweyl}.
In this geometry, a general scale-invariant
field equation and a geodesic equation have been obtained first by~\mbox{\citet{Dirac73}} and~\citet{Canu77}. Noticeably, in this context, it is possible to have
scale invariant field and geodesic equations in the presence of a non-zero cosmological 
constant.

At this stage, we do not know whether, in her wisdom, Nature presents the properties of
gauge invariance simultaneously with a non-zero cosmological constant.
However,  our role as scientists is to explore new paths, particularly in the present context
where 95\% of the matter--energy in the Universe is in an unknown form.
The observations are the judge; they will tell whether this is true or wrong. In~this context, there already are some encouraging 
first signs as shown below.

A limited form of scale invariance is also present in MOND,
which is remarkably successful in accounting for the high observed rotation velocities of spiral galaxies.
MOND implies a dilatation invariance where lengths 
and time are multiplied by a constant $\lambda$-factor, 
\begin{equation}
(r', t' ) \; \rightarrow \; (\lambda r, \lambda t)\, ,
\label{mondsc}
\end{equation}
\noindent
where $\lambda$ is independent from time.
Expressions~(\ref{ds}) and~(\ref{mondsc}) are in fact similar for comparisons concerning the present epoch; however,
the results are not necessarily the same, when cumulative effects over the ages intervene.

The scale-invariant vacuum (SIV) theory presents several promising features.
The cosmological equations predict an acceleration of the cosmic expansion~\citep{Maeder17a}, 
and the test analyses performed are in agreement with observations.
The growth of the density fluctuations appears fast enough during the matter era 
so that it is not necessary to call for dark matter~\citep{MaedGueor19}. Other tests of the SIV model are performed
(related to the dynamics in galaxies based on the weak field approximation) 
concerning the mass excess in clusters of galaxies, 
the growth with age of the ``vertical'' velocity dispersion in the Milky Way
\citep{Maeder17c}, and the rotation velocities of galaxies~\citep{MaedGueor20}.

Section~\ref{basic} summarizes the basic properties of the SIV theory and gives the cotensorial form of the equations.
Section~\ref{models} examines the cosmological equations, their results, and properties.
The relation between $H_0$ and the age of the Universe is studied in Section~\ref{age}.
Three other basic cosmological tests are analyzed in Section~\ref{mz}: the distances,
the magnitude--redshift $m-z$ or Hubble diagram is studied 
with the SNIa results of the joint analysis of the SDSS-II and SNLS observations by~\citet{Betou14},
and with a recent quasar and GRB data by~\citet{Lusso19}, 
the past expansion rates $H(z)$ analyzed by~\citet{Jesus18b} and 
finally a study of the predictions of the redshift drifts.
Section~\ref{weakf} summarizes some basic properties and tests of the weak field approximation (Newton like),
including in Section~\ref{drho} results on the growth of density fluctuations in the early Universe. 
Section~\ref{concl} gives the conclusions and further perspectives. 

\section{The Basic Equations and Properties of the SIV Theory} \label{basic}

\vspace{6pt}
\subsection{Integrable Weyl Geometry, Cotensor Expressions, and the Scale Invariant Field Equations} \label{weyl}

The SIV theory is based on the Integrable Weyl Geometry (IWG), some properties of which are given in Appendix~\ref{hweyl}.
In addition to a metric form $ds^2= g_{\mu \nu} dx^{\mu} dx^{\nu} $, this geometry is endowed with a scalar field $\Phi$:
\begin{equation}
\Phi \, = \, \ln \lambda \, , \quad \mathrm{with} \quad \kappa_{\nu} \, \equiv \, -\Phi_{,\nu} =
- \frac{\partial \ln \lambda}{\partial x^{\nu}} \, ,
\label{ka}
\end{equation}
\noindent
where $\lambda$ is the scale factor, expressing a transformation of the line element as in Equation~(\ref{ds}).
The~quantity $\kappa_{\nu}$ is the {\it{metrical connexion}}\footnote{We use the French spelling for metric connection 
to emphasize its relation to is construction within the Integrable Weyl Geometry (IWG) framework 
since within the standard Einstein GR such fields are zero by construction.}, 
it is a fundamental parameter of the IWG, where the lengths can undergo gauge changes of the form,
\begin{equation}
\ell'\,=\,\lambda(x^{\nu})\,\ell\,. \label{length}
\end{equation}

The scale factor $\lambda$ could in principle depend on the 4--coordinates $x^{\nu}$, 
but, because of homogeneity and isotropy, we consider it to be a function of time only. 
The quantities with a prime refer to Riemann Geometry of GR, while the quantities without 
a prime refer to the IWG. The integrable Weyl's space is conformally equivalent to
a Riemann space (defined by $g'_{\mu\nu}$)
via the $\lambda$-mapping that represents gauge~re-scaling: 
\begin{equation}
g'_{\mu\nu}\,=\,\lambda^{2}\,g_{\mu\nu}\,.\label{conformal}
\end{equation}

In the space with a prime (GR), the metrical connexion $\kappa'_{\nu}=0 $.
Such conformal mappings have been related to space-time deformations
and have been studied in connection with a large class of Extended
Theories of Gravity~\citep{Capozziello07,Capozziello11}. 
According to~\citet{Capozziello07}, the space-time deformations act like a
force that deviates the test particles from the unperturbed motions.
These~authors also express the corresponding geodesics and general
field equations that are consistent with those by 
\citet{Canu77} and~\citet{BouvierM78} as shown below.
{In relation to other works in the field, we may mention that Meierovich 
\cite{Meierovich12} is using a longitudinal vector field (cf. the derivative in Equation (\ref{ka}) )
to represent in the framework of GR the effects of the dark components of the Universe. 
The acceleration is a quadratic function of this vector field there, 
while in our SIV context it is a linear~function.
}

The cotensor analysis has been developed~\citep{Weyl23,Eddington23,Dirac73,Canu77}, 
with some further complements about the derivation of the equation of geodesics from an action principle, 
the study of the notion of parallel displacement, of isometries, and Killing vectors~\citep{BouvierM78}.
Standard scalars, vectors, or tensors that transform by following the rule: 
\begin{equation}
Y'^{\, \nu }_{\mu} \, = \, \lambda^{n} \, Y^{\nu}_{\mu} 
\label{co}
\end{equation}
upon a gauge change~(\ref{length}) consistent with the conformal transformation (\ref{conformal}), 
respectively define co-scalars, co-vectors, or~co-tensors of power $n=\Pi(Y^{\nu}_{\mu} )$. 
If $n=0$, one has an in-scalar, in-vector or in-tensor, such
objects are invariant upon a scale transformation.
In general, the derivative of a scale-covariant object (co-tensor, co-vector or co-scalar) is not scale covariant; 
furthermore, the derivative of a scale-invariant object is not scale-invariant either. 
However, one can write a modified scale-covariant derivative of an object, which has the properties of 
scale covariance, respectively of scale invariance when $n=0$. 
Such scale covariant derivatives are called ``co-covariant derivatives'' by~\citet{Dirac73}.
The co-covariant derivatives have the same power as the original object. 
For example, the co-covariant derivative (marked by an asterisk $_{*}$) of a coscalar S of power $n=\Pi(S)$ is
\begin{equation}
S_{* \mu} \, = \, \partial_{\mu} S - n \kappa_{\mu} S \,.
\end{equation}

The scale factor $\lambda$ is considered to be a co-scalar of power $\Pi(\lambda)=-1$ by imposing the requirement 
$\lambda_{*\mu}=\lambda_{,\mu}-\Pi(\lambda)\lambda\kappa_{\mu}\equiv 0 \Leftrightarrow$ Equation~(\ref{ka}),
which assures the commutativity of the scaling~(\ref{co}) and the co-covariant derivative operation. 
Furthermore, $\kappa_{\mu}$ is then a zero power $\Pi(\kappa_{\mu})=0$ gauge vector 
($\kappa_{\mu}\rightarrow\kappa'_{\nu}=\kappa_{\nu}+\partial_{\nu}\ln\lambda$) assuring that the 
co-covariant derivatives preserve the power of the object they are applied on. 
For a covector of power $n=\Pi(A)$, the co-covariant derivatives are: 
\begin{eqnarray}
A_{\mu*\nu}\, & = & \,\partial_{\nu}A_{\mu}-^{*}\Gamma_{\mu\nu}^{\alpha}A_{\alpha}-n\kappa_{\nu}A_{\mu},\,\\{}
A_{*\nu}^{\mu}\, & = & \,\partial_{\nu}A^{\mu}+^{*}\Gamma_{\nu\alpha}^{\mu}A^{\alpha}-n\kappa_{\nu}A^{\mu},\,\label{eq:co-cov_der}\\
\mathrm{with}\quad^{*}\Gamma_{\mu\nu}^{\alpha} & = & \Gamma_{\mu\nu}^{\alpha}+g_{\mu\nu}\kappa^{\alpha}\,-g_{\mu}^{\alpha}\kappa_{\nu}-g_{\nu}^{\alpha}\kappa_{\mu}.
\label{eq:co-Chr_sym}
\end{eqnarray}

Here, $^{*}\Gamma_{\mu\nu}^{\alpha}$ is a modified Christoffel symbol,
while $\Gamma_{\mu\nu}^{\alpha}$ is the usual Christoffel symbol.

The first derivatives of a co-tensor have the following expressions:
\begin{eqnarray}
T^{\mu \nu}_{* \rho} =T^{\mu \nu}_{, \rho}+^{*}\Gamma_{\rho\sigma}^{\mu} T^{\sigma \nu}+
^{*}\Gamma_{\rho\sigma}^{\nu} T^{\mu\sigma}-n \kappa_{\rho}T^{\mu \nu} \; ,\\
T_{\mu\nu * \rho} =T_{\mu\nu , \rho}- ^{*}\Gamma_{\mu \rho}^{\sigma} T_{\sigma \nu}-
^{*}\Gamma_{\nu\rho}^{\sigma} T_{\mu\sigma}-n \kappa_{\rho}T_{\mu \nu} \; .
\end{eqnarray}

Second and higher co-covariant derivatives of scalars, vectors, and tensors can also be expressed.
Pursuing the developments, a corresponding Riemann--Christoffel tensor with all symmetry properties can be obtained;
see, for example,~\citet{Dirac73}. The contracted Riemann--Christoffel tensor appears to be an intensor; it writes
\begin{equation}
R^{\nu}_{\mu} = R'^{\nu}_{\mu} - \kappa^{; \nu}_{\mu} - \kappa^{ \nu}_{;\mu}
- g^{\nu}_{\mu}\kappa^{ \alpha}_{;\alpha} -2 \kappa_{\mu} \kappa^{\nu}
+ 2 g^{\nu}_{\mu}\kappa^{ \alpha} \kappa_{ \alpha} \, ,
\label{RC}
\end{equation}
where $R'^{\nu}_{\mu}$ is the usual expression.
The sign ``;'' indicates a covariant derivative with respect to the mentioned coordinate. 
The total curvature $R$ in the scale-invariant context is :
\begin{equation}
R \, = \, R' -6 \kappa^{\alpha}_{; \alpha}+6 \kappa^{\alpha} \kappa_{\alpha}, \ 
\label{RRR}
\end{equation}
\noindent
where $R'$ is the total curvature in a standard Riemann geometry. We note that, in all expressions, 
the additional terms depend on $\kappa_\nu$.
For more details on the cotensor calculus, the interested reader may
read Chapter VII of {\it{``The Mathematical Theory of Relativity''}} by Eddington~\citep{Eddington23},
as well as~\citet{Dirac73}. We also point out that~\citet{Canu77}
provided a short summary of the cotensor analysis.

The above expressions, developed in the IWG, consistently lead to a general scale invariant field equation 
\citep {Canu77} that contains additional terms depending on $\kappa_{\nu}$, 
\begin{eqnarray}
R'_{\mu \nu} - \frac{1}{2} \ g_{\mu \nu} R' - \kappa_{\mu ;\nu} - \kappa_{ \nu ;\mu}
-2 \kappa_{\mu} \kappa_ {\nu} 
+ 2 g_{\mu \nu} \kappa^{ \alpha}_{;\alpha}
- g_{\mu \nu}\kappa^{ \alpha} \kappa_{ \alpha} =
-8 \pi G T_{\mu \nu} - \lambda^2 \Lambda_{\mathrm{E}} \, g_{\mu \nu}, \, 
\label{field}
\end{eqnarray}
\noindent
where $G$ is the gravitational constant, taken here as a true constant.
The cosmological term in Einstein's GR-equation is $\Lambda_{\mathrm{E}} g'_{\mu \nu}$, which according
to expression~(\ref{conformal}) transforms like
\begin{equation}
\Lambda_{\mathrm{E}} g'_{\mu \nu} \, = \, \Lambda_{\mathrm{E}} \lambda^{2}\,g_{\mu\nu}\, \equiv \, \Lambda \, g_{\mu \nu} \, .
\label{Lambda}
\end{equation}

Here, $\Lambda= \lambda^2 \Lambda_{\mathrm{E}}$ in the SIV context; it is a coscalar of power $\Pi(\Lambda)=-2$. 
Thus, the field equation in the above form~(\ref{field}) consistently is a scale invariant expression.

The second member of the RHS of~(\ref{field}) is scale invariant, and so is the first one.
This means that the energy-momentum tensor behaves like an in-scalar, if $G$ is considered to be in-scalar as well,
\begin{equation}
T_{\mu \nu} \, = \,T '_{\mu \nu} \, .
\label{tmn}
\end{equation}

This expression has some implications on the behavior of the relevant densities.
The tensor $T_{\mu \nu}$ being scale invariant, one may write
$
( p+\varrho) u_{\mu} u_{\nu} -g_{\mu \nu } p =
( p'+\varrho') u'_{\mu} u'_{\nu} -g'_{\mu \nu } p' \, .
$
The velocities $u'^{\mu}$ and $u'_{\mu}$ transform as follows:
\begin{eqnarray}
u'^{\mu}&=&\frac{dx^{\mu}}{ds'}=\lambda^{-1} \frac{dx^{\mu}}{ds}= \lambda^{-1} u^{\mu} \, , \nonumber \\
\mathrm{and} \; \;
u'_{\mu}&=&g'_{\mu \nu} u'^{\nu}=\lambda^2 g_{\mu \nu} \lambda^{-1} u^{\nu} = \lambda \, u_{\mu} \, .
\label{pl1}
\end{eqnarray}

Thus, the energy-momentum tensor transforms by 
$$( p+\varrho) u_{\mu} u_{\nu} -g_{\mu \nu } p =
( p'+\varrho') \lambda^2 u_{\mu} u_{\nu} - \lambda^2 g_{\mu \nu } p' \, .$$
This implies the following scaling of $p$ and $\varrho$~\citep{Canu77},
\begin{equation}
p = p' \, \lambda^2 \, \quad \mathrm{and} \quad \varrho = \varrho' \, \lambda^2 \, .
\label{ro2}
\end{equation}

Thus, pressure and density are not scale invariant but are coscalars of power $\Pi(\rho)=-2$,
like the term $\Lambda= \lambda^2 \Lambda_{\mathrm{E}}$.
However, to avoid any ambiguity, we keep all expressions with $\Lambda_{\mathrm{E}}$,
the Einstein cosmological constant.

\subsection{The General Field Equation from an Action Principle and the Lagrangian} \label{Lagrangian}

The generalization of the Einstein GR-field equations to~(\ref{field}) and
the relationships of the Riemann curvature and Ricci tensor to their
counterparts within the IWG could be derived via the substitution of~(\ref{conformal}) 
directly into the corresponding tensors or via variation of the action:
\begin{equation}
\int\lambda^{2}\left(*R\right)\sqrt{g} \,d^{4}x \, .
\end{equation}

Here, we use an asterisk in front of $R$ to emphasize that $(*R)$ is a IWG co-scalar of non-zero power
and therefore some additional objects are needed to make the whole action in-scalar object.
The~in-scalar properties of this action are seen easily by re-calling
that $\lambda$ is a co-scalar of power $\Pi(\lambda)=-1$,
while $\left(*R\right)$ is of power $\Pi(*R)=-2$ since it is a contraction of the in-tensor $R_{\mu\nu}$
using $g^{\mu\nu}$, $\Pi(g_{\mu\nu})=2$, and $g=\det(g_{\mu\nu})$
is therefore a co-scalar of power $\Pi(g)=8$~\citep{Canu77}. 
The corresponding variational equations are the same as those derived by substitution
of~(\ref{conformal}) directly into the corresponding tensors. However, the variational
approach allows for an extension of the Lagrangian to include additional
terms in $\lambda$ and its co-tensor derivatives ($*\mu$). 
That is, a term $c_{1}\lambda^{*\mu}\lambda_{*\mu}$ and a term $c_2\lambda^{4}$. 
The $c_{1}$ term is irrelevant because $\lambda_{*\mu}=\lambda_{,\mu}-\varPi(\lambda)\lambda \kappa_{\mu}=0$
due to the definition of the metrical connexion $\kappa_{\mu}=-\partial_{\mu}\ln(\lambda)$.
The quartic term then is related to the cosmological constant upon
the substitution: $\Lambda\sim c_2\lambda^{2}$. It has been shown by
\citet{Canu77} that the equations corresponding to the variation with
respect to the scale factor $\lambda$ do not produce new field equations
but can be derived via trace of the corresponding Einstein field equations. 
In our case,~(\ref{field}) results in:
\begin{equation}
R'+6\kappa_{\mu}\kappa^{\mu}-6\kappa_{\:;\mu}^{\mu}=
4\Lambda+8\pi G T_{\mu}^{\mu}. 
\label{effCC}
\end{equation}

The requirement of scale-invariance of the action leads to 
$\lambda{\cal T}_{\mu}^{\mu}=\lambda GT_{\:\mu}^{\mu}=\frac{\delta{\cal L}}{\delta\lambda}$
\mbox{(see~\citet{Canu77})}. Thus, in the absence of matter (${\cal L}=0$
$\Rightarrow$ ${\cal T}_{\mu}^{\mu}=0$) and in the Einstein gauge ($\lambda=1$), the
cosmological constant is related to the Ricci scalar $R'=4\Lambda_{\mathrm{E}}$;
however, in a flat IWG vacuum ($R'=0$), one then obtains an equation about the
scaling factor: 
\begin{equation}
6 \, \kappa_{\mu} \, \kappa^{\mu}-6\, \kappa_{\:;\mu}^{\mu} \, = \, 4 \, \Lambda \,.
\label{LagCC}
\end{equation}

In general, if $T_{\mu}^{\mu}$ is not zero, it could contribute to an effective cosmological constant.
In what follows, a specific gauge is chosen which allows for utilization of the 
flat IWG vacuum as a foundation for the scale invariant vacuum theory.

\subsection{Fixing the Gauge}

As it stands, the field Equation~(\ref{field}) is undetermined due to
the gauge symmetry of the equations. The same problem appears in General Relativity, where
the under-determinacy of GR is resolved by the choice of coordinate
conditions. Here, one needs to impose some gauging conditions to define
the scale factor $\lambda$. 
The key hypothesis is made that 
{\it{the properties of the empty space, at macroscopic scales, are scale invariant}}
as in~\citet{Maeder17a}.
This choice is justified since the usual equation of state for the
vacuum $p_{\mathrm{vac}}=-\varrho_{\mathrm{vac}}$ is precisely the
relationship permitting $\varrho_{\mathrm{vac}}$ to remain
constant for an adiabatic expansion or contraction~\citep{Carr92}.
Under the above key hypothesis, 
one is left with the following condition for empty ``vacuum'' spacetime
deduced from the field Equation~(\ref{field}): 
\begin{equation}
\kappa_{\mu;\nu}+\kappa_{\nu;\mu}+2\kappa_{\mu}\kappa_{\nu}
-2g_{\mu\nu}\kappa_{;\alpha}^{\alpha}+g_{\mu\nu}\kappa^{\alpha}\kappa_{\alpha}
=\lambda^{2}\Lambda_{\mathrm{E}}\,g_{\mu\nu}.\label{fcourt}
\end{equation}

The geometrical terms $R'_{\mu\nu}$ and $R'$ of the field equation have disappeared from~(\ref{field}), 
since the de Sitter metric for an empty space endowed with a cosmological constant 
is conformal to the Minkowski metric where $R'_{\mu\nu}=0=R'$~\cite{deSitter1917}. 
In particular, if $3 \lambda^{-2}/(\Lambda_{\mathrm{E}} \tau^2)=1$,
where the time parameter $\tau$ is discussed in the 4th remark after 
Equation~(\ref{diff1}) and in more detail in~(\ref{tau}) and in~\cite{Maeder17a}.
This condition is noticeably satisfied by the solution of~(\ref{fcourt}), 
as shown below, if $\lambda$ is a function of time.

Since $\lambda$ may only be a function of time $t$,
only the zero component of $\kappa_\mu$ is non-vanishing. 
Thus,~the coefficient of metrical connexion becomes
\begin{eqnarray}
\kappa_{\mu ;\nu} = \kappa_{0 ;0} = \partial_0 \kappa_0 = \frac{d \kappa_0}{dt} \equiv
{\dot{\kappa}_0} \, .
\label{k0}
\end{eqnarray}

We use the Minkowski metric with a positive time-component signature,
having checked that it is compatible with the scale invariant field equation in empty space~\citep{Maeder17a}.
However, the results that follow are the equivalent if one is to use mostly positive metric signature.
The 0 and the 1, 2, 3 components of what remains from the field Equation~(\ref{fcourt}) become respectively:
\begin{equation}
3 \kappa^2_0 \, = \,\lambda^2 \, \Lambda_{\mathrm{E}} \, ,
\label{k1}
\end{equation}
\begin{equation}
2 \dot{\kappa}_0 - \kappa_0^2 = -\lambda^2 \Lambda_{\mathrm{E}} \, .
\label{k2}
\end{equation}

Notice that~(\ref{k1}) and~(\ref{k2}) correspond to the two non-trivial coordinate equations from 
(\ref{fcourt}); in this respect, the trace of~(\ref{fcourt}), given also by~(\ref{LagCC}), can be obtained
by multiplying the spatial coordinate Equation~(\ref{k2}) by $(-3)$ and, when added to~(\ref{k1}), one recovers 
(\ref{LagCC}) for the choice of gauge~(\ref{k0}) .
From~(\ref{ka}), one has $\kappa_0 \, = - \dot{\lambda}/\lambda$ (using $c=1$ units),
while~(\ref{k1}) and~(\ref{k2}) lead to $\dot{\kappa}_0 \, = - \kappa_0^2$ along with:
\begin{eqnarray}
\Lambda_{\mathrm{E}} \, = 3 \, \frac{ \dot{\lambda}^2}{\lambda^4} \, 
\quad \mathrm{and} \quad \Lambda_{\mathrm{E}}= 2 \, \frac{\ddot{\lambda}}{\lambda^3}
- \frac{ \dot{\lambda}^2}{\lambda^4} \, .
\label{diff1}
\end{eqnarray}

These expressions, which relate $\Lambda_{\mathrm{E}}$ and the scale factor $\lambda(t)$, 
result from our key hypothesis and some important remarks can be made about them:
\begin{itemize}
\item Since $\Lambda_{\mathrm{E}}$ represents the energy of the vacuum, it also establishes a relation between the energy density
of the vacuum and the scale factor $\lambda$ and its possible variations.
\item As noted by~\citet{Durrer08}, ``Instead of a cosmological constant, one may also introduce a scalar field or some other contribution to the energy-momentum tensor...'' , this is just what is happening here. The field $\Phi$ as defined in Equation~(\ref{ka})
is therefore a scalar field associated with the vacuum properties at macroscopic scales in the Universe.
\item In GR, $\Lambda_{\mathrm{E}}$ and therefore the properties of the empty space associated with it 
are considered to not depend on the matter content of the Universe. 
This is evident upon considering the Einstein gauge ($\lambda=1$) in~(\ref{effCC}) 
along with the fact that radiation does not contribute to the trace ${\cal T}_{\mu}^{\mu}$ and
the matter contribution vanishes since $\varrho_m\simeq a^{-3}\rightarrow 0$ in the limit of ever increasing scale factor $a$.
Thus, in standard GR, one is forced to accept $4\Lambda_{\mathrm{E}}=R'$. 
We adopt the same assumption here. This~means that the above expressions of $\Lambda_{\mathrm{E}}$~(\ref{diff1}) are valid
whatever the matter content of the Universe. 
In particular, if the IWG with $\lambda$ satisfying~(\ref{diff1}) is adopted, then the corresponding equations~(\ref{field})
would imply local no-cosmological constant Einstein GR equations for the matter fields (see~(\ref{E1}) and~(\ref{E2})).
\item 
The solution of the above two equations~(\ref{diff1}) is of the form:
$\lambda \, = \, \frac{y}{(t-b)^n} $.
Both Equations~(\ref{diff1}) imply $n=1$, while $y \, = \, \sqrt{\frac{3}{\Lambda_{\mathrm{E}}}}$ follows from the first equation. 
Any value of the parameter $b$ would satisfy the equations. However, the value of $b$ has to be consistent 
with the cosmological equations for the expansion factor $a(t)$ when applied to the considered case of an empty space.
The results below and in~\citet{Maeder17a} show that $b=0$ has to be taken in this case.
\item
The expression of $\lambda = \sqrt{\frac{3}{\Lambda_{\mathrm{E}}}} \, \frac{1}{ct}$ 
in current units (in calculations $c=1$ is usually taken). The~value $\lambda_0$ 
at the present time is then $\lambda_0= \sqrt{\frac{3}{\Lambda_{\mathrm{E}}}}$ (with $c=1$).
\end{itemize}

\section{Accelerating Cosmological Models} \label{models}

The main cosmological consequence of the scale invariant vacuum theory is that the models predict an acceleration
of the expansion as a result of the scalar field resulting from the properties of the vacuum.

\subsection{The Basic Equations of SIV Cosmology}

The application of the FLWR metric, appropriate for an homogeneous and isotropic space, to~the general field Equation~(\ref{field})
leads to some additional terms resulting from the $\kappa$-terms in the cosmological equations as obtained by~\citet{Canu77},
\begin{equation}
\frac{8 \, \pi G \varrho }{3} = \frac{k}{a^2}+
\frac{\dot{a}^2}{a^2}+ 2 \, \frac{\dot{\lambda} \, \dot{a}}{\lambda \, a}+
\frac{\dot{\lambda}^2}{\lambda^2} - \frac {\Lambda_{\mathrm{E}} \lambda^2}{3} \,
\label{E1p}
\end{equation}
\begin{equation}
-8 \, \pi G p = \frac{k}{a^2}+ 2 \frac{\ddot{a}}{a} + 2 \frac{\ddot{\lambda}}{\lambda}+\frac{\dot{a}}{a}^2
+ 4 \frac{\dot{a} \, \dot{\lambda}}{a \, \lambda}-\frac{\dot{\lambda^2}}{\lambda^2} -\Lambda_{\mathrm{E}} \, \lambda^2 \, .
\label{E2p}
\end{equation}

These equations do not have known analytical solutions for the expansion factor $a(t)$ and 
have not led to satisfactory cosmological models. However, if we accept 
``the postulate of GR that gravitation couples universally to all energy and momentum" 
\citep{Carr92}, we may account for expressions~(\ref{diff1}) 
characterizing the properties of {\it{a homogeneous and isotropic}}  empty space.
Thus, with the two relations in~(\ref{diff1}), the equations obtained from~(\ref{field}) nicely simplify to~\citep{Maeder17a}
\begin{equation}
\frac{8 \, \pi G \varrho }{3} = \frac{k}{a^2}+\frac{\dot{a}^2}{a^2}+ 2 \,\frac{\dot{a} \dot{\lambda}}{a \lambda} \, ,
\label{E1}
\end{equation} 
\begin{equation}
-8 \, \pi G p = \frac{k}{a^2}+ 2 \frac{\ddot{a}}{a}+\frac{\dot{a^2}}{a^2}
+ 4 \frac{\dot{a} \dot{\lambda}}{a \lambda} \, .
\label{E2}
\end{equation}

The combination of these two equations leads to
\begin{equation}
-\frac{4 \, \pi G}{3} \, (3p +\varrho) = \frac{\ddot{a}}{a} + \frac{\dot{a} \dot{\lambda}}{a \lambda} \, .
\label{E3}
\end{equation}

Term $k$ is the curvature parameter which takes values $0$ or $\pm 1$; 
$p$ and $\varrho$ are the pressure and density in the IWG.
Einstein cosmological constant has disappeared due to the account of Equations~(\ref{diff1}).
These three equations differ from the classical ones, in each case, only by one additional 
term containing $\frac{\dot{a} \, \dot{\lambda}}{a \, \lambda} $, which depends on time $t$.
If $\lambda(t)$ is a constant, 
one gets the usual equations. 
Thus, at any fixed time, the effects that do not depend on time evolution are just those 
predicted by GR (for example, the gravitational shift in stellar spectral lines).
Significant departures from GR may essentially appear in cosmological evolution over the ages. 

The significance of the additional term appearing in~(\ref{E3}), 
where we have $\dot{\lambda}/\lambda \, = \, - \frac{1}{t}$, is as follows.
If there is an expansion with $\dot{a} > 0$, it represents 
{\it{ an acceleration in the direction of the expansion and opposed to gravitation}}. 
On the contrary, if there is a contraction, it is 
{\it{ an acceleration in the direction of the contraction, thus adding to gravitation}}.
Equations~(\ref{E1}) to~(\ref{E3}) are fundamentally different from those of the 
$\Lambda$CDM models: a variable term replaces the 
cosmological constant $\Lambda_{\mathrm{E}}$. 
The new term implies an acceleration (variable with time) of the expansion or contraction.
In this connection, we note that the need to have a time-dependent term
has been emphasized by several authors,
\citet{Sahni14}, \mbox{\citet{Sola15}},~\citet{Ding15}, and~\citet{Sola16}.

\subsection{Density, Geometry Parameters, and Conservation Laws} \label{basicprop}

Let us examine some of the general properties of the above equations 
in the matter dominated era with $p=0$.
Equation~(\ref{E1}) divided by $H = \frac{\dot{a}}{a}$ can be written, 
\begin{equation}
\frac{8 \, \pi G \varrho }{3 \,H^2} = 1 + \frac{k}{a^2 \, H^2} - \frac{2}{ H \, t} \, .
\label{E11}
\end{equation}

With the following definitions, considered at time $t_0$,
\begin{equation}
\Omega_{\mathrm{k}} \equiv -\frac{k}{a^2 H^2} \, \quad
\Omega_{\lambda} \equiv \frac{2}{ H \, t} \, \quad 
\Omega_{\mathrm{m}} \equiv \frac{\varrho}{\varrho_{\mathrm{c }}}\, \quad 
\mathrm{with} \quad \varrho_{\mathrm{c }} \equiv \frac{ 3 \, H^2}{8 \pi G} \, ,
\label{defomega}
\end{equation}
we get
\begin{equation}
\Omega_{\mathrm{m}} \, + \, \Omega_{\mathrm{k}} \, + \Omega_{\lambda} = \, 1 \, .
\label{Omegasum}
\end{equation}

The term $\Omega_{\lambda}$ is associated with the scale factor $\lambda(t)$.
{{Let us examine how this acceleration term behaves over the ages. 
Based on the model results of Equation~(\ref{tin}),
where, in the case of a flat model, we have $H(t) \, = \, {2 \, t^2}/{(t^3 - \Omega_{\mathrm{m}})}$;
this implies the following behavior of $\Omega_{\lambda}$:
\begin{equation}
\Omega_{\lambda} \, = \, \frac{2}{ H \, t} \, = \,\frac{t^3-\Omega_{\mathrm{m}}}{t^3} \, = 
\,1- \frac{\Omega_{\mathrm{m}}}{t^3} \, . \label{omll} 
\end{equation} 

We see that relation~(\ref{Omegasum}) is valid at the present time, while at other epochs  
it becomes of the form~(\ref{omll}), the two forms being evidently compatible at $t_0=1$. 
Thus, as time is going on, $\Omega_{\lambda}$ is increasing, while the mass density term is  
decreasing. This shows that the acceleration term is gaining importance over the ages. 
Conversely in the early stages, the role of $\Omega_{\lambda}$ was smaller 
compared to that of $\Omega_{\mathrm{m}}$---notice that the first expression in Equation~(\ref{tin}) implies 
$\Omega_{\lambda}=0$ at $t_{in}$ when the scale factor takes the value $a(t_{in})=0$. 
This is also confirmed by the SIV solutions for the relativistic era~\cite{Maeder19}, 
which shows a braking of the expansion  with $a(t) \, \sim \, t^{1/2}$ in the radiative era.  
}}

Here, $\varrho_{\mathrm{c}}$
is the usual value of the {{present}} critical density in Friedman models, and so does $\Omega_{\mathrm{m}}$.
(We make this choice in order to not introduce new confusing definitions; 
note that another definition of the critical density could be introduced in the SIV cosmology~\citep{Maeder17a}). 
While in Friedman's models there is only one model density corresponding to $k=0$, 
here, for $k=0$, there is a variety of models with different $\Omega_{\mathrm{m}}$ and $\Omega_{\lambda}$. 
For all models, whatever the $k$-value, the density parameter $\Omega_{\mathrm{m}}$ remains smaller than 1.
For $k=-1$ and $k=0$, this is clear since $2 \, /(t_0 H_0)$ is always positive for an expanding Universe.
For $k=+1$ (negative $\Omega_{\mathrm{k}}$), numerical models confirm that 
$ \Omega_{\mathrm{k}} \, + \Omega_{\lambda} $ is always positive so that $\Omega_{\mathrm{m}} < 1$. Thus,
scale invariant models only exist for $\Omega_{\mathrm{m}} \, < \, 1$. In other words, scale invariance is limited to 
densities lower than critical. That is, the presence of matter in the Universe with a density higher than
the critical one is killing the scale invariance. 

Let us examine the geometry parameter $q_0 = - \frac{\ddot{a}_0 \, a_0}{\dot{a}^2_0}$.
Equation~(\ref{E2}), in the absence of pressure and 
divided by $H^2_0$, becomes
\begin{equation}
-2 \, q_0 +1 - \Omega_{\mathrm{k}}= \frac{4}{H_0 t_0} \, .
\label{qk}
\end{equation}

For flat models with $k=0$, we get

\begin{equation}
q_0 \,= \frac{1}{2} -\frac{\Omega_{\mathrm{k}}}{2} -\Omega_{\lambda} = \frac{1}{2} -\Omega_{\lambda} \, 
\label{qzerom}
\end{equation}

This establishes a relation between the deceleration parameter $q_0$ and the matter content for the
scale invariant cosmology. Using~(\ref{Omegasum}), we also have 
\begin{equation}
q_{0} = \frac{1}{2} -\Omega_{\lambda} \, = \, \Omega_{\mathrm{m}} - \frac{1}{2} \, ,
\label{qkk}
\end{equation}
which provides a very simple relation between basic parameters. 
For $\Omega_{\mathrm{m}}=0.30$, 0.20, or 0.10, we would
get $q_0=-0.20\, -0.30$ or $-0.40$.
The above basic relations are different from those of the $\Lambda$CDM, which is expected 
since the basic Equations~(\ref{E1})--(\ref{E3}) are different. 
Let us recall that, in the $\Lambda$CDM model with $k=0$, one has
$q_0\, = \, \frac{1}{2}\Omega_{\mathrm{m}} -\Omega_{\Lambda}$. For 
$\Omega_{\Lambda}=0.70$, 0.80 or 0.90 and 
$\Omega_{\mathrm{m}}=0.30$, 0.20 or 0.10, one gets 
$q_0=-0.55$, --0.70, or --0.85, 
which shows that the second derivative of the expansion factor $a(t)$ behaves differently with
the density in the two kinds of models.

Conservation laws and invariances are intimately connected, 
thus the scale invariance of the empty space must influence the conservation laws. 
Let us use the above cosmological equations for that. 
We first rewrite~(\ref{E1}) as follows and take its derivative,
\begin{equation}
8 \, \pi G \varrho\, a^3 = 3\, k\, a+ 3 \, \dot{a}^2 a + 6 \, \frac{\dot{\lambda}} {\lambda} \dot{a}\, a^2 \, , 
\end{equation}
\begin{eqnarray}
\frac{d}{dt} (8 \, \pi G \varrho \, a^3 ) 
= -3 \dot{a}\, a^2 \, 
\left[-\frac{k}{a^2}- \frac{\dot{a}^2}{a^2}-2 \frac{\ddot{a}}{a} 
-2 \frac{\ddot{a}}{ \dot{a}} \frac{\dot{\lambda}}{\lambda}-2 \frac{\ddot{\lambda}}{\lambda}
- 4 \frac{\dot{a}\, \dot{\lambda}}{a \, \lambda} +2 \frac{\dot{\lambda}^2}{\lambda^2}\right] \, ,
\label{derE1}
\end{eqnarray}

Equations~(\ref{E1}),~(\ref{E3}), and~(\ref{diff1}) lead to welcome simplifications,
\begin{eqnarray}
\frac{d}{dt} (8 \, \pi G \varrho\, a^3 ) \, 
=-3\, \dot{a}\, a^2 \left[8 \, \pi G p+\frac{{a}}{ \dot{a}} \frac{\dot{\lambda}}{\lambda}
\left(8 \, \pi G \, p+ \frac{8 \, \pi G \varrho}{3}\right) \right] \, .
\end{eqnarray}

This can also be written
\begin{equation}
d(\varrho a^3) + 3\,p \, a^2 da+3 \, p \, a^3 \frac{d \lambda}{\lambda}+ a^3 \varrho \frac{d \lambda}{\lambda}= 0 \,.
\end{equation}
\begin{equation}
\mathrm{and} \; \; \; 3 \, \frac{da}{a} + \frac{d \varrho}{\varrho}+\frac{d \lambda}{\lambda}+ 3 \, \frac{p}{\varrho}\; \frac{da}{a}+
3 \, \frac{p}{\varrho} \; \frac{d\lambda}{\lambda} = 0 \, .
\label{conserv1}
\end{equation}

We may also express it in a form similar to the usual conservation law,
\begin{equation}
\frac{d(\varrho a^3)}{da} + 3 \, p \,a^2+ (\varrho+3\, p) \frac{a^3}{\lambda} \frac{d \lambda}{da} = 0 \, .
\label{conserv2}
\end{equation}

These last two equations give the law of conservation of mass-energy in the scale invariant cosmology.
For a constant $\lambda$, we evidently recognize the usual conservation law.
We now write the equation of state in the general form,
$p \, = \, w \, \varrho $, with $c^2$ =1,
where $w$ is taken here as a constant.
The equation of conservation~(\ref{conserv1}) becomes
$
3 \, \frac{da}{a} + \frac{d \varrho}{\varrho}+\frac{d \lambda}{\lambda}+ 3 \, w \, \frac{da}{a}+
3 \, w \, \frac{d\lambda}{\lambda} = 0$,
with the following simple~integral,

\begin{equation}
\varrho \, a^{3(w+1)} \, \lambda ^{(3w+1)} \,= const. 
\label{3w}
\end{equation}

Notice that the integer powers of $a$ and $\lambda$ correspond to familiar values in the literature:
such as radiation, matter, and vacuum energy $w=1/3,0,-1$. 
The case $w=-1$ is the case of a cosmological constant $\varrho_\Lambda\propto\Lambda_E\lambda^2$. Notice that such scaling of the cosmological constant is consistent with scale-invariant stress energy tensor~(\ref{tmn})
of an ideal fluid discussed earlier~(\ref{ro2}); that is, a scale-invariant vacuum energy.
For radiation, $w=1/3$, the corresponding trace of the stress-energy tensor is zero 
$T_{\mu}^{\mu}=T_{\mu\nu}g^{\mu\nu}=\varrho(1-3w)=0$; therefore, 
it has no contribution to a potential effective cosmological constant via~(\ref{effCC}).
For $w=0$, this is the case of ordinary matter of density 
$\varrho_{\mathrm{m}}$ without pressure, 
\begin{equation}
\varrho_{\mathrm{m}} \, a^3 \, \lambda =const. 
\label{consm}
\end{equation}
\noindent
which means that the inertial and gravitational properties (respecting the Equivalence principle) within a covolume should slowly 
increase over the ages with $M=M'/\lambda$.
However, interestingly enough, this implies that the gravitational potential $ \phi = G \, M/r$ of a given object 
is an inscalar, meaning that it stays constant with time, since $r$ also behaves like $1/\lambda$. 
A change of the inertial and gravitational mass is not a new fact; it is well known in Special Relativity, where 
the effective masses change as a function of their velocity. In the standard model of particle physics, the constant masses of elementary
particles originate from the interaction of the Higgs field~\citep{Higgs14,Englert14} in the 
vacuum with originally massless particles. 
In addition, in the $\Lambda$CDM, the resulting acceleration of a gravitational system 
does not keep the total energy of the system unchanged~\citep{Krizek16}. 
The assumption of scale invariance of the vacuum \mbox{(at large scales)} 
makes the inertial and gravitational masses to slowly slip over the ages; 
however, it is by a rather limited amount in realistic cosmological models. For example, in a flat model with 
$\Omega_{\mathrm{m}}=0.30$, $\lambda$ varies only from 1.4938, 
at the origin (the Big-Bang where $a=0$) to 1 at present.
The~lower the density, the bigger the variations.

\subsection{Results of Cosmological Models} \label{num}

We study here the SIV models for the case of ordinary matter with $w=0$, for a flat Universe with $k=0$ as supported
by the~\citet{Planck14}. The numerical solutions for $k= 0, \pm 1$ were obtained by~\citet{Maeder17a}.
From~(\ref{E1}), we have the basic equation in the case $k=0$ considered here,
\begin{equation}
\dot{a}^2 a \, t - 2 \, \dot{a}\, a^2 + k \, a\, t - C \, t^2 =0 \, , \quad \mathrm{with}
\quad C=\frac{8 \, \pi G \varrho_{\mathrm{m}} \, a^3 \lambda}{3} \, .
\label{E12}
\end{equation}

Time $t$ is expressed in units of $t_0=1$, at which we also
assume $a_0 = 1$. The origin, the Big-Bang, occurs when $a(t)=0$ at an initial time $t_{\mathrm{in}}$ 
(see Equation~(\ref{tin})). The constant $C$ contains the scale factor $\lambda$. 
With  $\lambda_0 =\sqrt{3/\Lambda_E}$ at the present time, it is 
$ C=\frac{8 \, \pi G \varrho_{\mathrm{m}}}{\sqrt{3 \, \Lambda_{\mathrm{E}}}}$. 
However, as we may see below, the exact value of $\lambda_0$ has no influence on the solution of the cosmological models.
Indeed, we notice~that, if we have a solution $a$ vs. $t$, then $(x \, a)$ vs. $(x \, t)$ is also a solution.
This demonstrates an overall scale invariance of~(\ref{E12}).
The above equation implies at the present time $t_0$,
\begin{equation}
H^2_0 \, - 2 \, H_0 \, - C \, = 0 \, , \quad \mathrm{leading \; to} \quad H_0= 1 \pm \sqrt{1+C} \, ,
\label{h2}
\end{equation}
where we will take the sign ``+'' since $H_0$ is positive.
Equation~(\ref{E1}) becomes~(\ref{E11}) with $\Omega_{\mathrm{m}} = \varrho/ \varrho_{\mathrm{c}}$ and gives:
\begin{equation}
H_0 \, = \, \frac{2}{1- \Omega_{\mathrm{m}}} \, \quad \mathrm{and \; with \;~(\ref{h2}) \; one\; has:} \quad
C \, = \frac{4 \, \Omega_{\mathrm{m}}}{(1- \Omega_{\mathrm{m}})^2} \, .
\label{C}
\end{equation}

\begin{figure}[H]
\centering
\includegraphics[width=.80\textwidth]{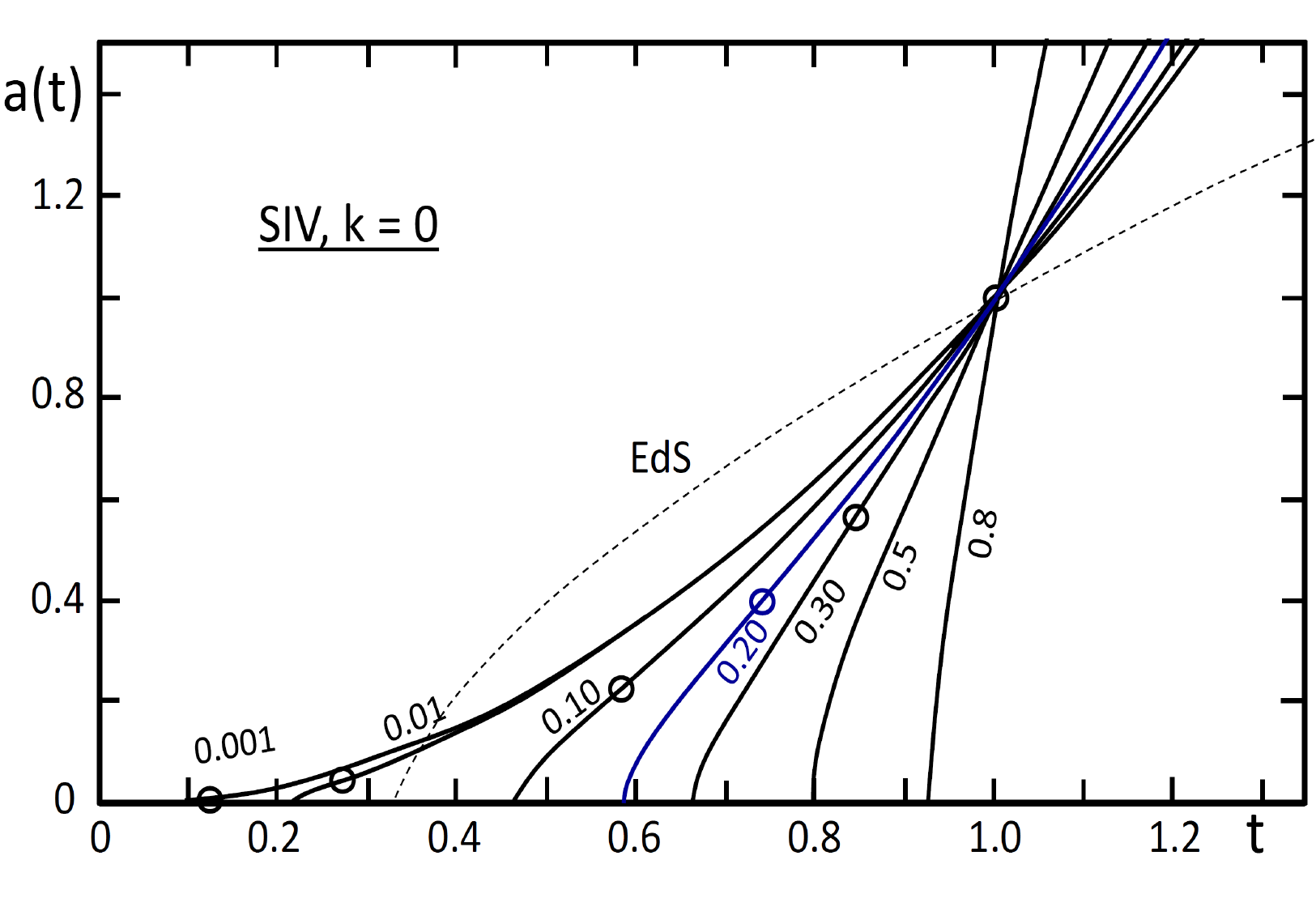}
\caption{Some solutions of a(t) for the models with $k=0$~\citep{Maeder17a}. 
The curves are labeled by the values of $\Omega_{\mathrm{m}}$. 
The Einstein--de Sitter model (EdS) is indicated by a dotted line. 
The small circles on the curves show the transition point between braking
$(q >0) $ and acceleration $(q<0)$. For $\Omega_{\mathrm{m}}=0.80$, this point is at $a= 2.52$. 
The blue curve corresponds to $\Omega_{\mathrm{m}}=0.20$.}
\label{atzero}
\end{figure}

For $\Omega_{\mathrm{m}}$ ranging from $0$ to $1$, $C$ varies between 0 to $\infty$, 
according to the above relation. The reason is that, when $\Omega_{\mathrm{m}}$ tends towards 1, 
the solution $a(t)$ tends towards a singularity: the curve $a(t)$ vs. $t$ is a straight vertical line tending to infinity, 
its derivative $H_0$ is infinite and the same for $C$ according to Equations~(\ref{h2}) and~(\ref{C}), 
which follows from the general Equation~(\ref{E1}). 
An alternative way to see this behavior for C, from its definition in Equation~(\ref{E12}), 
is to recall that  the general solution of the Equations~(\ref{diff1}) 
has the form $\lambda \, = \, {\lambda_0}/{(t-t_{in})} $ which diverges when $t$ tends towards~$t_{in}$.
The general trend of $a(t)$ is well illustrated in Figure~\ref{atzero}. 
The integration of the differential Equation~(\ref{E12}) leads to a useful analytical relation given by~\citet{Jesus18}
for flat space ($k=0$):
\begin{equation}
a(t) \, = \, \left[\frac{t^3 -\Omega_{\mathrm{m}}}{1 - \Omega_{\mathrm{m}}} \right]
\label{R}
\end{equation}

One can see again that relation~(\ref{C}) between the constant $C$ and $\Omega_{\mathrm{m}}$
does not depend on the exact form of $\lambda_0$ and so does the solution~(\ref{R}) for $a(t)$.
The initial time of the Big-Bang is given by $a(t_{\mathrm{in}}) = 0$, 
and the Hubble expansion rate by $H(t) = \dot{a}/a$,
\begin{equation}
t_{\mathrm{in}} \, = \, t_0 \, \Omega^{1/3}_{\mathrm{m}} \, , \quad \mathrm{and} \quad 
H(t) \, = \, \frac{2 \, t^2}{t^3 - \Omega_{\mathrm{m}}} \, .
\label{tin}
\end{equation}

For $H_0$, this is evidently consistent with expression~(\ref{C}). All the above values are on a timescale where $t_0=1$.
We have taken as a boundary condition $a(t)=0$. Some models with a vector field adopt $da/dt=0$~\cite{Meierovich12},
which then leads to a regular initial evolution instead of a singular Big-Bang. If one would adopt such an initial condition, 
Equation~(\ref{E1}) would lead to a static empty Universe for $k=0$. 
Thus, it seems difficult to adopt a different initial condition than $a(t)=0$.
Achieving a regular initial evolution via $da/dt=0$ may not be needed  for $k=0$ since 
the Big-Bang could be viewed as an apparent singularity within the SIV paradigm. This point has been discussed 
towards the end of the section {\it ``Fixing the gauge within the scale-invariant vacuum theory"} in~\cite{MaedGueor19}. 
Here, we recall briefly that upon the use of a logarithmic time  $t\rightarrow\tau=\ln(t/t_0)$
the ``now'' is at $\tau=0$, the~Big Bang is at $- \infty $, upon cosmic time when $t_{in}=0$,
and the spacetime is locally Lorentz invariant since the metric tensor is now Minkowski like. 

\begin{figure}[H]
\centering
\includegraphics[width=.70\textwidth]{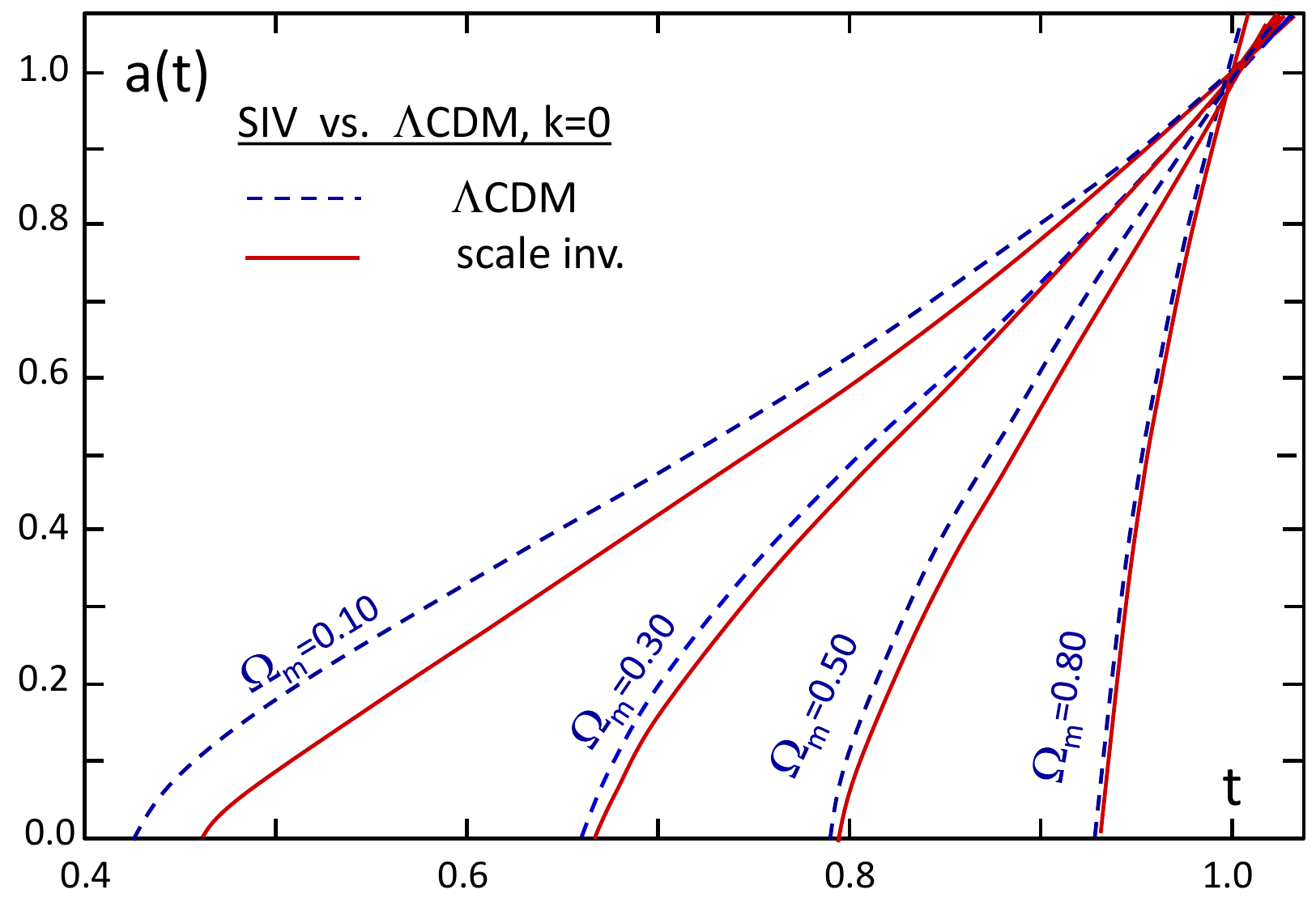}
\caption{Comparisons of the $a(t)$ functions of the $\Lambda$CDM and scale invariant models for given values of
$\Omega_{\mathrm{m}}$~\citep{Maeder17a}.}
\label{ascaLCDM}
\end{figure}

Figure {\ref{atzero}} shows the model results for various values of $\Omega_{\mathrm{m}}$.
After an initial phase of braking, there is an accelerated expansion, which goes on all the way. 
Only the empty model starts with a horizontal tangent; in this case, the effects of scale invariance are the largest.
All other models with matter start explosively 
with very high values of $H={\dot{a}} /a$ and a positive value of $q$, indicating braking, before turning later to an
accelerated expansion (at points marked by small circles in Figure~\ref{atzero}).
Figure~\ref{ascaLCDM} shows the comparison of the scale invariant and
$\Lambda$CDM models both with $k=0$. The curves 
of both kinds of models are similar with larger differences 
for lower density parameters. 
These results imply that for most observational tests the predictions of the scale invariant 
models are not far from those of the $\Lambda$CDM models.

The ratio ${\lambda (t)}/{\sqrt{3 \, \Lambda_{\mathrm{E}}}}$ varies between 
$1/t_{\mathrm{in}}= 1/(t_0 \Omega^{1/3}_{\mathrm{m}})$ and 1 at the present time.
For an empty space, this ratio varies enormously, between $\infty$ at the origin and 1 at present.
As soon as matter is present, the range of the above factor falls dramatically.
For $\Omega_{\mathrm{m}}= 0.30$, ${\lambda (t)}/{\sqrt{3 \, \Lambda_{\mathrm{E}}}}$ 
varies only from 1.4938 to 1.0 between the Big-Bang and now. Thus, the 
presence of less than 1 H-atom per cubic meter
is sufficient to drastically reduce the amplitude of the domain of $\lambda$-variations.
For $\Omega_{\mathrm{m}}$ approaching 1,
the effects of scale invariance disappear and the cosmological solutions tend towards those of GR. 
Indeed, we have seen that there is no scale invariant solution for $\Omega_{\mathrm{m}}> 1$.
Thus, in the line of the remark by~\citet{Feynman63} in the Introduction,
we see that the presence of even very tiny amounts of matter in the Universe very rapidly 
tend to kill scale invariance. The point is that, for realistic values of $\Omega_{\mathrm{m}}$, 
the effects appear to be not yet completely killed and may thus deserve the present investigation.
The elliptic and hyperbolic scale invariant models have been discussed in~\citet{Maeder17a}.

On the whole, it is a satisfactory result that the scale invariant models, which are resting on a very simple hypothesis,
consistently predict an accelerated expansion. We now turn to comparisons with observations, which are the judge of the models.

\section{Relation between the Hubble Constant \boldmath{$H_0$} and the Age of the Universe in the SIV Cosmology} \label{age}

\vspace{6pt}
\subsection{Theoretical Predictions}

The relation between the Hubble constant $H_0$ and the estimated age $\tau_0$ of the Universe 
is a fundamental test. For a given curvature parameter $k$, it only depends on 
$\Omega_ {\mathrm{m}}$. The test is simple in its principle, but not necessarily in practice due
to the tension in the observational determinations of $H_0$, as well as uncertainties in the age of
the Universe.

Let us call $\tau$ the age in the usual scale of years, or seconds, while $t$ is a time in the scale where
the present age $t_0=1$, $t_{\mathrm{in}}$ is the initial time in this timescale. We{ can define $\tau$ as
\begin{equation}
\tau \, = \, \frac{t- t_{\mathrm{in}}}{1- t_{\mathrm{in}}} \Delta_0 \, ,
\label{tau}
\end{equation}
where $\Delta_0$ is the present age of the Universe in the chosen units, typically $\Delta_0 = 13.8 \cdot 10^9$ yr~\citep{Frie08}.
The~Hubble expansion rate in the $t$ and $\tau$ scales are 
related by 
\begin{equation}
H(\tau) \, = \, \frac{1}{a} \, \frac{da}{dt} \frac{dt}{d\tau} \, = \, H(t) \, \frac{dt}{d\tau} \, ,
\end{equation}
\begin{equation}
\mathrm{with} \quad d\tau= \frac{dt}{1- t_{\mathrm{in}}} \Delta_0 \,= \frac{dt}{1- \Omega^{1/3}_{\mathrm{m}}} \Delta_0 \, .
\end{equation}

Thus, for $k=0$, the Hubble function $H(\tau)$ becomes:
\begin{equation}
H (\tau) \, = 
\, \frac{2 \, t^2}{t^3 - \Omega_{\mathrm{m}}} (1- \Omega^{1/3}_{\mathrm{m}}) \frac{1}{\Delta_0} \,
\label{htau}
\end{equation}
\begin{equation}
\mathrm{with} \quad H_0 = \, \frac{2 (1- \Omega^{1/3}_{\mathrm{m}}) }{1 -\Omega_{\mathrm{m}}} \frac{1}{\Delta_0} \,.
\label{Ho}
\end{equation}

Concerning the units, let us note that a recession velocity of $70$ km s$^{-1}\, $ Mpc$^{-1}$ corresponds to
$(1/\Delta_0) =2.2685 \cdot 10^{-18}\,$ s$^{-1}$, i.e., $\Delta_0 = 4.408 \cdot 10^{17}$ s or 13.969 Gyr.
This correspondence gives easy calculations of $H_0$ vs. age in their proper units.
Figure~\ref{Hovsage} illustrates with black lines the values of $H_0$ in km s$^{-1}\, $ Mpc$^{-1}$ vs. the density parameter
$\Omega_{\mathrm{m}}$ as predicted by Equation~(\ref{Ho}) for three different values of the age of the Universe: 
$\Delta_0 =12.6$, 13.8, and 15 Gyr.

\begin{figure}[H]
\centering
\includegraphics[width=.75\textwidth]{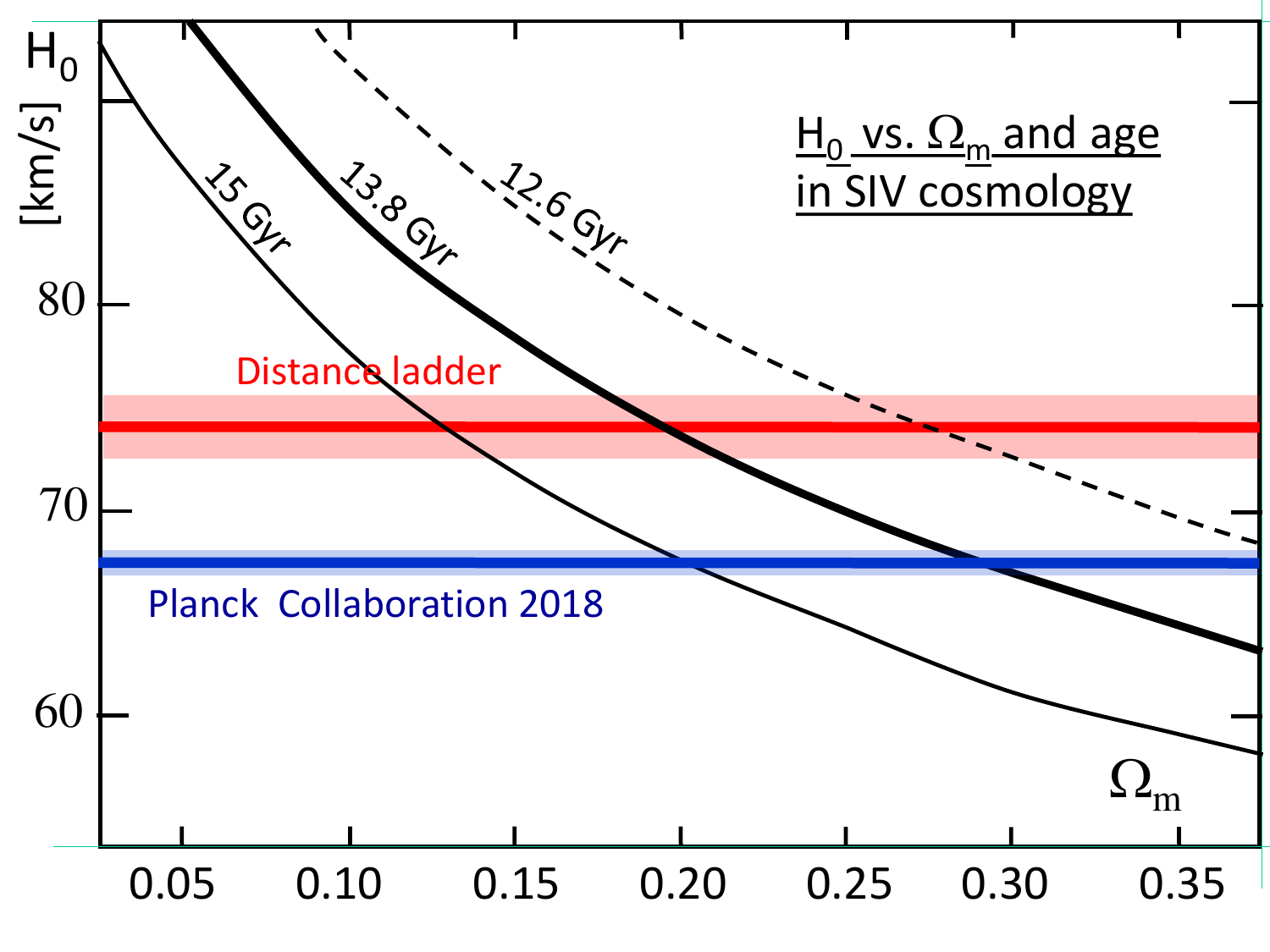}
\caption{The relation between the Hubble constant $H_0$ and $\Omega_{\mathrm{m}}$ 
predicted by the SIV theory (Equation~(\ref{Ho})) for three different
values of the age of the Universe. The result from the distance ladder with
a new calibration of the LMC Cepheids of 74.03 $(\pm 1.42)$ km s$^{-1}\, $ Mpc$^{-1}$ by~\mbox{\citet{Riess19}} is indicated by 
a thick red line, with its uncertainty. The thick blue line indicates the value of 67.4~$(\pm 0.5)$~km~s$^{-1}\, $ Mpc$^{-1}$
obtained by the~\mbox{\citet{Planck18}}.}
\label{Hovsage}
\end{figure}

\subsection{Comparisons with Observational Data}

The Saga of the determinations $H_0$ and of the age of the Universe could fill several books.
Let us be short and first consider $H_0$, always expressed below in km s$^{-1}\, $ Mpc$^{-1}$.
There is a bi-polarization around two values: 
a high value from the cosmic ladder formed mainly by Cepheids
and Ia supernovae and a low value from the CMB data obtained by the Planck collaboration. On the high side, 
\mbox{\citet{Frie08}} gave a value $H_0= 72 \pm 5 $ (in the same units),
\citet{Free10} obtained $H_0 = 73 \pm 2$ (random) $ \pm 4$ (systematic), 
a value later supported by~\citet{Riess16} from HST data for SNIa which lead to $H_0 = 73 \pm 1.75$. 
Recently, with a new anchor based on LMC Cepheids for the SNIa ladder,~\citet{Riess19} obtained a slightly larger value
$H_0 =74.03 \pm 1.42$; this is the value we retain in Figure~\ref{Hovsage} for $H_0$ based on the distance ladder.
We note that some higher values have been given recently.~\citet{Tully16} obtained 
values of $H_0 = 75 (\pm 2)$ or 76.2 from various distance estimators.
From the mass distribution and time delay of a strongly lensed quasar,~\citet{Shajib19}
have given 74.2 (+2.7$-$3.0). From distant lensed galaxies,~\citet{Chen19} found a value of 76.8 $(\pm 2.6)$.
Let us turn to the lower values, based on CMB analysis and BAO oscillations. The~\citet{Planck14} 
gives $H_0 = 67.3 \pm 1.2$ within the six-parameter $\Lambda$CDM cosmology. This result is supported by
further studies by the same team 
\citep{Planck18} obtaining $H_0 = 67.4 \pm 0.3$. The combination of BAO and SN Ia data into an inverse distance ladder
leads to a value $H_0 = 67.3 \pm 1.1$~\citep{Aubourg15}, rather similar results were obtained by
\citet{L'Hui17} with $H_0=68.49 \pm 1.53$. 

Let us now turn to the estimates of the age of the Universe, independent of the cosmological models. 
The age estimates from globular clusters range between 12 and 15 Gyr, with a general reference age of
13.8 Gy~\citep{Frie08}, which we consider here. However, some interesting claims for a possible higher age
have been made.~\citet{VdB14} studied several old halo subgiants
with accurate parallaxes and high quality spectra and obtained a value of 14.3 Gyr for HD 140283; adding 0.4 Gyr for the time interval
between the Big-Bang and the first star formation would lead to an
age of 14.7 Gyr. Such an age of about one Gyr greater than the 
standard reference value is also supported by~\mbox{\citet{Rouk18}}, on the basis on Galactic Bulge microlensed low-mass stars.

Figure~\ref{Hovsage} compares the various data and SIV models. Clearly, the reference age together with distance ladder $H_0$
support a value of $\Omega_{\mathrm{m}} \simeq 0.20$. A higher age of, say, 14.7 Gyr, would push $\Omega_{\mathrm{m}}$
down to about 0.14. The above age and $H_0$ does not depend on any cosmological model; this is not the case for
CMB and BAO data. Even if they do not explicitly rely on the $\Lambda$CDM models, these determinations assume
that the oscillations responsible for the CMB peaks and BAO oscillations obey a gravitation law, which is different from what is
assumed by the SIV theory, as given by the field and geodesic Equations~(\ref{field}) and~(\ref{geod}). 
Thus, in the present context, a determination based on this data, despite their high accuracy, would unfortunately 
not be a theoretically consistent process.

On the whole, the $H_0$ vs. age test supports a density parameter $\Omega_{\mathrm{m}} \simeq 0.20$, maybe down to 0.14.
It~is not surprising to find a value lower than the current $\Lambda$CDM value of about 0.30, 
since, in the context of the scale invariant theory, the assumption of dark matter is not a necessary one. However, we note that the 
final word concerning the value of $\Omega_{\mathrm{m}}$ could only be given by models of the cosmological nucleosynthesis 
in the appropriate context.

\section{Basic Cosmological Tests of the SIV Cosmology} \label{mz}

\vspace{6pt}
\subsection{Distances} \label{ddist}

Distances intervene in many cosmological tests. 
The distance of an object with coordinate $r_1$~($\vartheta_1=0, \varphi_1=0$)
depends on the evolution of the expansion factor $a(t)$.
There are several definitions of distances: the proper motion distance $d_{\mathrm{M}}$, the angular diameter distance
$d_{\mathrm{A}}$, and the luminosity distance $d_{\mathrm{L}}$,
\begin{equation}
d_{\mathrm{L}} \,= \, (1+z) d_{\mathrm{M}} \,= \, (1+z)^2d_{\mathrm{A}} \, , \quad \mathrm{with}
\quad d_{\mathrm{M}} \, = \,a_0 \, r_1 \, ,
\label{distances}
\end{equation}

The relations between these three distances are model independent and $d_{\mathrm{M}}$ is: 
\begin{equation}
a_0 \, r_1 = c \int^z_0 \frac{dz}{H(z)} \, ,
\label{pmd}
\end{equation}
\noindent
where $c$ is the speed of light in vacuum.
One can express $H$ as a function of $z$ by using Equation~(\ref{R}),
$(t^3-\Omega_{\mathrm{m}})=(1-\Omega_{\mathrm{m}}) \, a^{3/2}$:
\begin{equation}
H(t) = \frac{2 \left[(1-\Omega_{\mathrm{m}}) a^{3/2}+\Omega_{\mathrm{m}}\right]^{2/3} }
{(1-\Omega_{\mathrm{m}}) a^{3/2}} \,
\end{equation}

With the expressions $H_0=2/(1-\Omega_{\mathrm{m}}) $, $a_0/a= 1+z$, and with $a_0=1$, one gets:
\begin{equation}
\frac{H(z)}{H_0} = \left(\Omega_{\mathrm{m}} a^{-9/4}+(1-\Omega_{\mathrm{m}}) a^{-3/4}\right)^{2/3}=
\left(\Omega_{\mathrm{m}} (1+z)^{9/4}+(1-\Omega_{\mathrm{m}}) (1+z)^{3/4}\right)^{2/3} \, ,
\label{hz}
\end{equation}
which can also be written as
\begin {equation}
H(z) \, = \, \frac{2 \, \sqrt{1+z}}{1- \Omega_{\mathrm{m}}} \, \left( \Omega_{\mathrm{m}}\,(1+z)^{3/2}+
(1- \Omega_{\mathrm{m}}) \right) ^{2/3} \, .
\label{hz2}
\end{equation}

Then, the proper motion distance is:
\begin{equation}
a_0 r_1 = \frac{c}{H_0} \int^{z_1}_0 \frac{dz}{ \left(\Omega_{\mathrm{m}} (1+z)^{9/4}+
(1-\Omega_{\mathrm{m}}) (1+z)^{3/4}\right)^{2/3} },\,
\label{pmd2}
\end{equation}
which can be integrated numerically.

Figure~\ref{distan} shows the angular distance $d_{\mathrm{A}}$ as a function of redshift for scale invariant models
of various $\Omega_{\mathrm{m}}$.
Up to redshift $z=2$, the SIV models give almost the same curve whatever 
$\Omega_{\mathrm{m}}$. At large $z$, the differences remain relatively limited, compared to the $\Lambda$CDM models.
This shows the need of accurate data
for determining of values $\Omega_{\mathrm{m}}$ and for discriminating
between the SIV and $\Lambda$CDM models~\citep{Maeder17a}.

\begin{figure}[H]
\centering
\includegraphics[width=.70\textwidth]{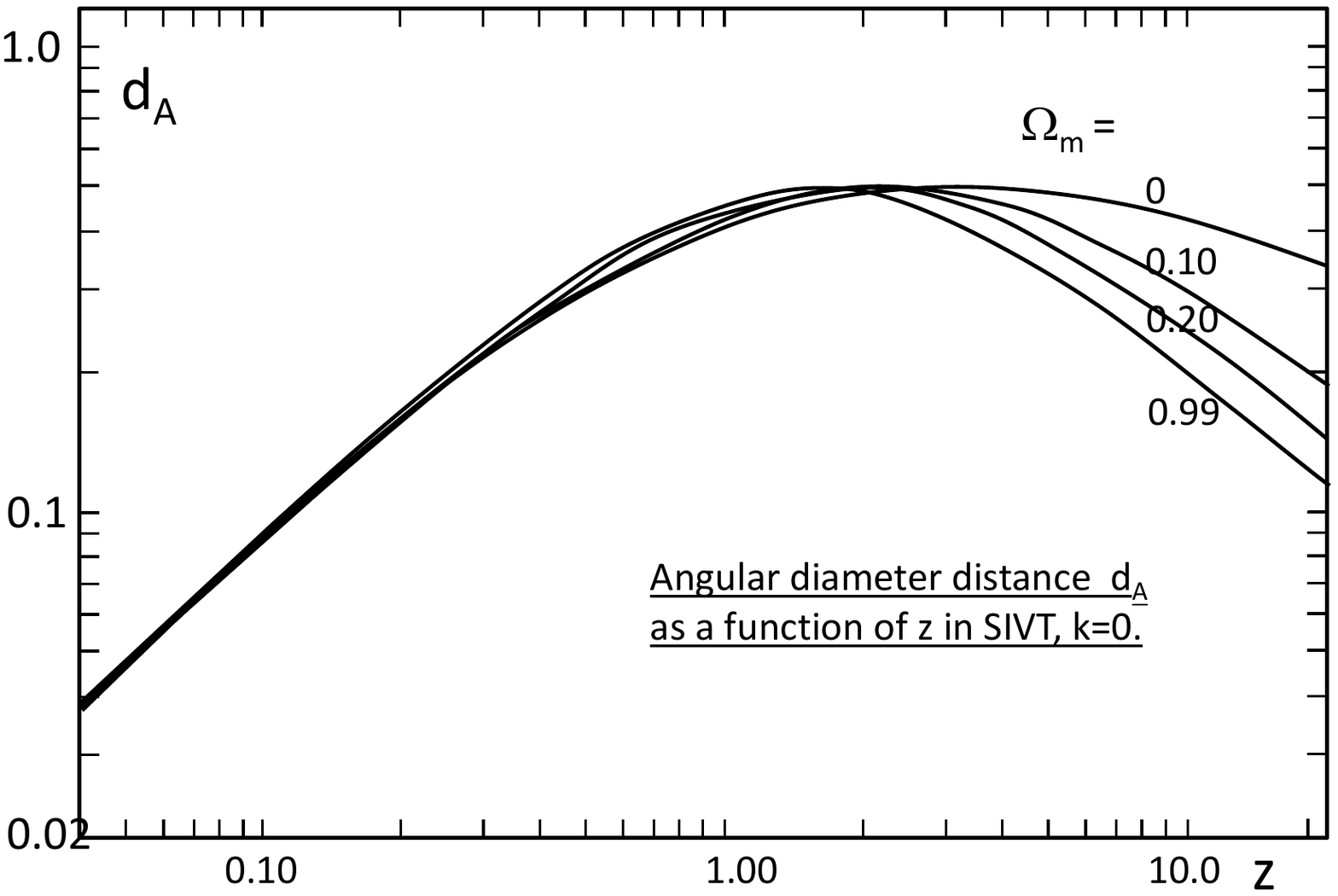} 
\caption{Comparison of the angular diameter distances for $\Lambda$CDM and SIV models of 
different density parameters $\Omega_{\mathrm{m}}$~\citep{Maeder17a}, with the scales $t_0=1, \, a_0=1$.
} 
\label{distan}
\end{figure}

\subsection{The Magnitude-Redshift Diagram of Distant Standard Candles}

The Hubble or magnitude-redshift diagram has been a major tool for the discovery of the accelerated expansion~\citep{Riess98,Perl99}. 
The flux $f$ received from a distant standard candle, like a supernova of type Ia, with a luminosity $L$
goes like $f = L/(4 \, \pi \, d^2_{\mathrm{L}})$ and the distance modulus $m-M$ of a source of coordinate $r_1$ is
\begin{equation}
m - M \, = \, const. + 5 \log (a_0 r_1) + 5 \log (1+z) \, , 
\end{equation}

\noindent
where the proper motion distance $a_0 r_1$ is obtained from~(\ref{pmd2}).
We choose the constant so that \mbox{$m-M = 33.22$} at redshift $z=0.01$.
Such calibration is based on the recent $(m-M)$ vs. $z$ diagram from 
the joint analysis of type Ia supernova observations obtained in the SDSS-II and SNLS collaboration
given in Figure 8 in ~\citet{Betou14}.
At this low redshift, the various models, whether $\Lambda$CDM or SIV, have a similar value of the above constant.
Relations $(m-M)$ vs. $z$ are calculated for a few scale invariant models for different density parameters
$\Omega_{\mathrm{m}}$. Table~\ref{table1} shows some results. 
We see that significantly different results between models of various 
$\Omega_{\mathrm{m}}$ only appear for redshift higher than~$z =1$.

\begin{table}[H] 
\vspace{-3mm} 
\caption{ Values of $(m-M)$ as a function of redshift $z$ for different density parameters $\Omega_{\mathrm{m}}$ }\vspace{-3mm}  
\label{table1}
\begin{center} 
\scriptsize
\begin{tabular}{cccccc}
\toprule
\boldmath{$z$} & \boldmath{$\Omega_{\mathrm{m}}=0.05$} & \boldmath{$\Omega_{\mathrm{m}}=0.10$} &\boldmath{$ \Omega_{\mathrm{m}}=0.20$} &
\boldmath{$\Omega_{\mathrm{m}}=0.30$} & \boldmath{$ \Omega_{\mathrm{m}}=0.40$ }\\
\midrule
0.10 & 38.35 & 38.35 & 38.34 & 38.32 & 38.31 \\
0.30 & 40.99 & 40.98 & 40.95 & 40.92 & 40.89 \\
0.70 & 43.23 & 43.20 & 43.13& 43.06 & 43.00 \\
1.00 & 44.24 & 44.19 & 44.09 & 44.00 & 43.92 \\
2.00 & 46.30 & 46.20 & 46.02 & 45.88 & 45.75\\
3.00 & 47.54 & 47.40 & 47.16 & 46.98 & 46.83 \\
4.00 & 48.43 & 48.24 & 47.97 & 47.76 & 47.58 \\
5.00 & 49.11 & 48.89 & 48.58 & 48.35 & 48.16 \\
6.00 & 49.66 & 49.42 & 49.08 & 48.83 & 48.63 \\
7.00 & 50.13 & 49.86 & 49.49 & 49.23 & 49.03 \\
8.00 & 50.52 & 50.23 & 49.85 & 49.58 & 49.37 \\
9.00 & 50.87 & 50.56 & 50.16 & 49.88 & 49.66 \\
10.0 & 51.18 & 50.85 & 50.44 & 50.15 & 49.93 \\
\bottomrule
\normalsize
\end{tabular}
\end{center}
\end{table}

\vspace{-8mm}

Figure~\ref{Betou} compares the SNIa results of the above-mentioned joint analysis of the 
SDSS-II and SNLS data~\citep{Betou14} with the scale invariant models. 
This analysis is based on a sample of 740 spectroscopically confirmed SNIa with high-quality light curves.
The sample benefits from a considerable improvement in the calibration of the luminosity scale. 
The best fit $\Lambda$CDM model represented by a black line is obtained for $\Omega_{\mathrm{m}}= 0.295 \pm 0.034$. 
The flat SIV models with $\Omega_{\mathrm{m}} = 0.10$ and 0.20 are indicated by red lines. 
We see, consistently with the plot of distances, that, for all considered redshifts, the differences between models
of various $\Omega_{\mathrm{m}}$ are very small. A scale invariant model with $\Omega_{\mathrm{m}}$ in the range 0.10--0.20 
encompasses the best fit $\Lambda$CDM model, thus suggesting a $\Omega_{\mathrm{m}}$ value of about 0.20, 
in good agreement with what we derived from the comparison of $H_0$ and the age of the Universe. 
Thus, the scale invariant models are consistent with the Hubble diagram for the available SNIa data, 
which show the acceleration of the expansion.
The major difference, however, is that the SIV models do not imply some unknown form of dark energy.

\begin{figure}[H]
\centering
\includegraphics[width=.85\textwidth]{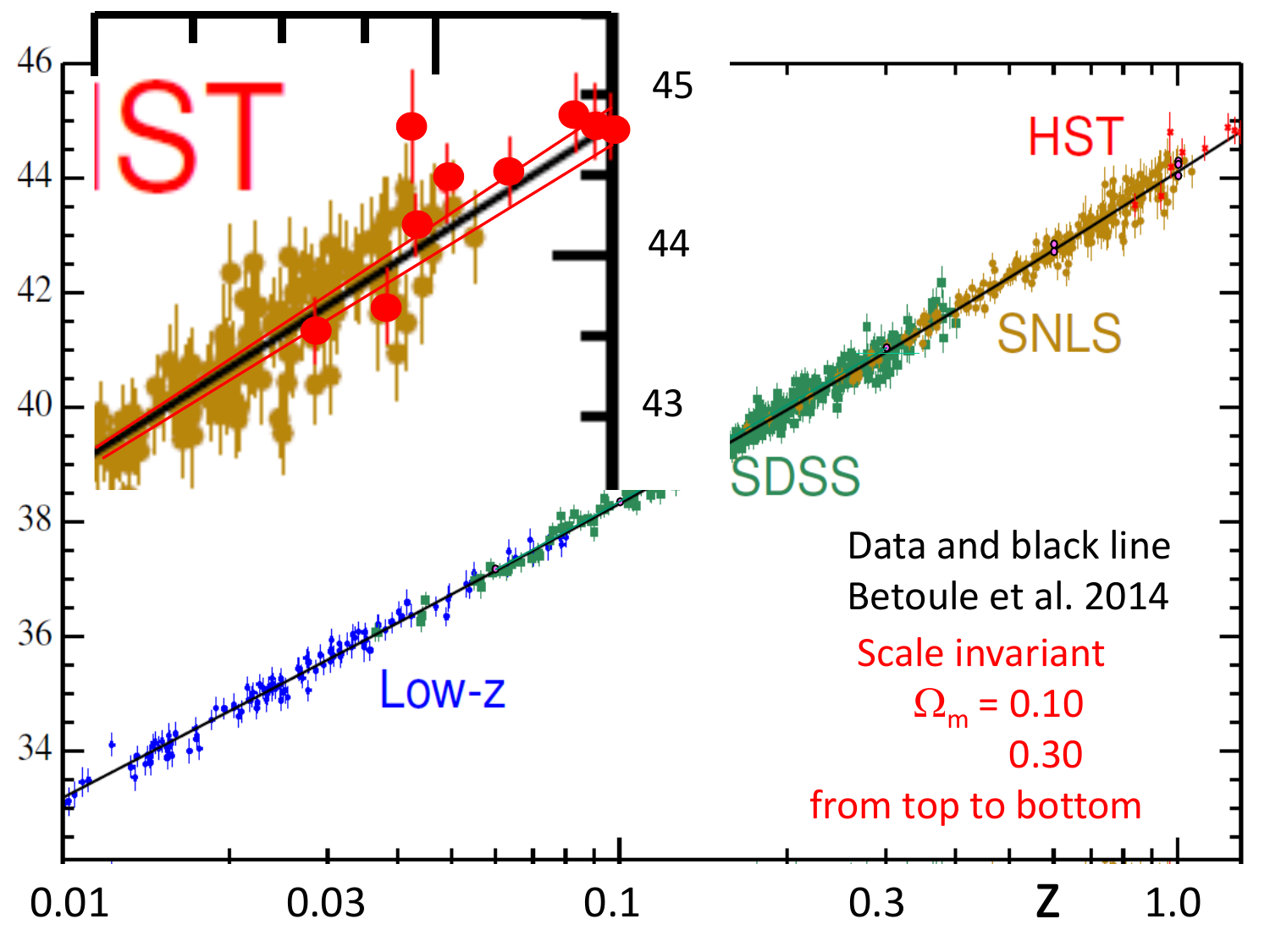} 
\caption{ The magnitude-redshift diagram by~\citet{Betou14} based on the joint analysis of SNIa 
observations in the SDSS-II and SNLS collaborations. The upper right corner of the plot is enlarged in an insert. 
The scale invariant models with $\Omega_{\mathrm{m}}=0.10$ and 0.30 are indicated by red lines. 
The black line is the best fit $\Lambda$CDM model for $\Omega_{\mathrm{m}}= 0.295$. 
}
\label{Betou}
\end{figure}

A new study of the Hubble diagram extended up to redshifts $z\simeq 7$ has been recently published by~\citet{Lusso19}. 
It is based on a sample of SNIa, quasars, and gamma-ray bursts (GRBs). The~SNIa sample
consists of 1048 supernovae (The Pantheon sample) of type Ia up to $z=2.26$ established by~\citet{Scolnic18} on the basis 
the PS1 Medium Deep Survey, together with HST samples and 
distances estimates from SDSS and SNL surveys. The reference of the magnitude scale used by Scolnic~et al. appears from
his plots to correspond to 33.17 at redshift $z=0.01$, instead of 33.22 
in the calibration by~\citet{Betou14}. This small difference is, however,
of little importance in the comparisons below.
The quasar
sample contains 1598 objects with high quality UV and X-ray
measurements for redshifts $z$ in the range of 0.04 to 5.1.
The distances are obtained from luminosity estimates
based on a nonlinear relation between the UV and X-ray flux
for the quasars~\citep{Risaliti19}. In Figure~\ref{Lusso}, the quasars have been binned in narrow ranges of redshifts; nevertheless, their
scatter remains important.

For the GRBs, the sample considered by Lusso et al. consists of a collection of data of GRBs 
established by~\citet{Demianski17a}.
For the distance determinations, the
relation between the peak photon energy and the isotropic equivalent 
radiated energy is applied according to the analysis by~\citet{Demianski17a}, who 
have shown that this relation appears to not depend on redshifts.
Quasars and especially GRBs offer a most remarkable extension of the Hubble
diagram to extreme distances, far beyond the range presently covered by SNIa data.

\begin{figure}[H]
\centering
\includegraphics[width=.85\textwidth]{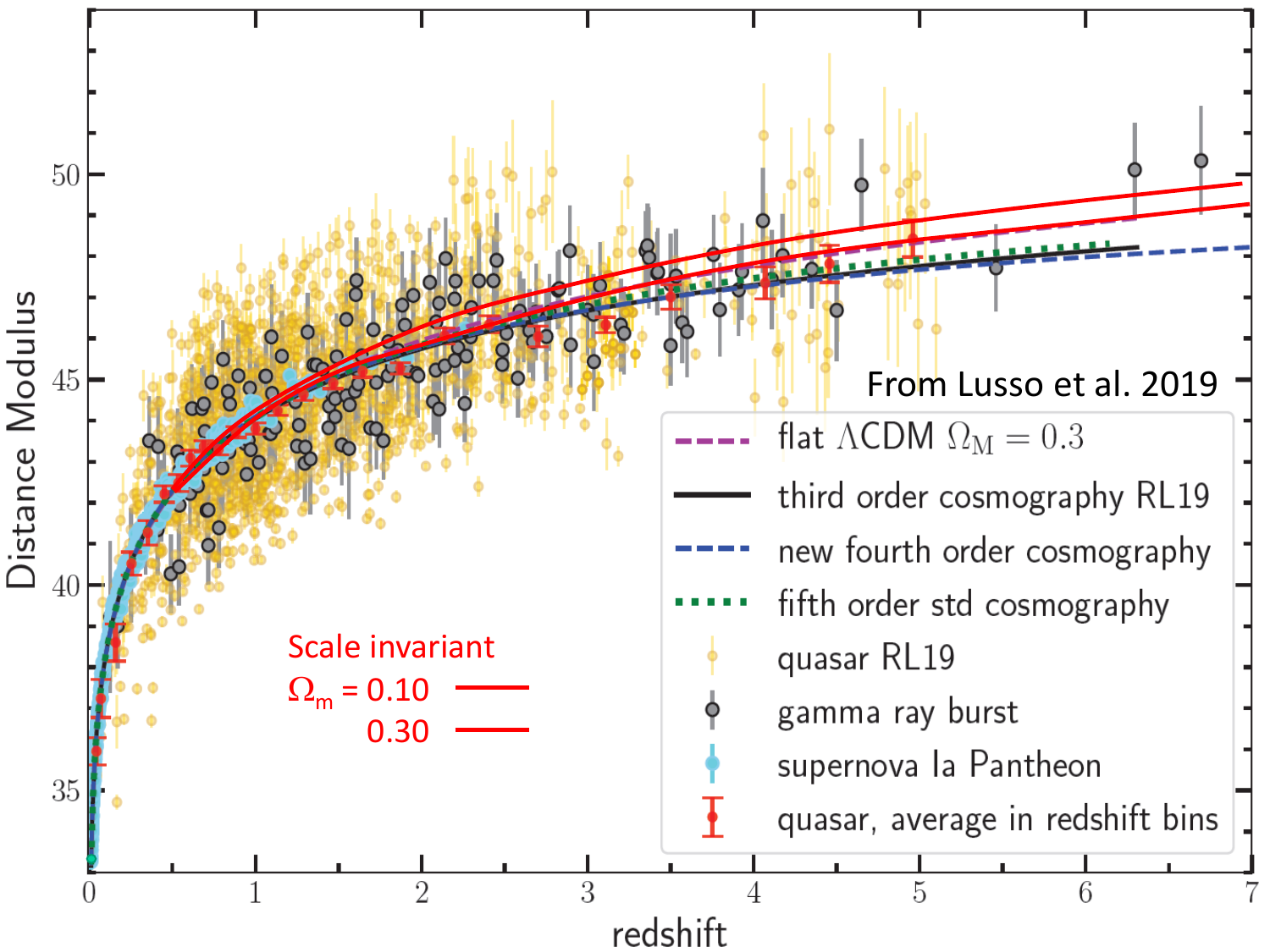}
\caption{The Hubble diagram for SNIa, quasars (binned), and GRBs from the samples collected by~\citet{Lusso19}. 
The various models considered by Lusso et al. are indicated. The two red lines show the flat scale invariant models
with $k=0$ and $\Omega_{\mathrm{m}}=0.10$ and 0.30. }
\label{Lusso}
\end{figure}

Figure~\ref{Lusso} shows the comparison of SNIa, quasar, and GRB data with various curves as indicated in the plot. The two red 
lines show the SIV models of Table~\ref{table1} for $\Omega_{\mathrm{m}}=0.10$ and 0.30. This last model
lies very close to the $\Lambda$CDM model with $ \Omega_{\mathrm{m}}=0.30$, which means that, for such
density parameters, the Hubble diagram will have difficulties to discriminate between the 
$\Lambda$CDM and SIV models. This also implies that the SIV models lie rather close to the
observations.

\citet{Lusso19} claim that there is a significant difference between
the $\Lambda$CDM models and the data, thus suggesting new physics.
These authors use the cosmographic method to express the luminosity distance as a function of redshift.
In this approach, which is model independent, the expansion factor $a(t)/a(t_0)$ is developed in a Taylor series of terms,
\begin{equation}
\frac{a(t)}{a(t_0)} = 1 +H_0(t-t_0)+\frac{ q_0}{2!}(t-t_0)^2+ \frac{j_0}{3!} (t-t_0)^3+
\frac{s_0}{4!} (t-t_0)^4 +\frac{l_0}{5!}(t-t_0)^5+{\it{O}}[(t-t_0)^6] \, .
\end{equation}

Then, the luminosity distance is also developed in a Taylor series as a function of $z$ up to $z^5$. 
The coefficients of
this development are themselves functions of the parameters $H_0, q_0, j_0, s_0, l_0$, the last three being called the jerk, snap, and lerk.
The coefficients of the development are all defined at $z=0$. 
We see from Figure~\ref{Lusso} that the third, fourth, and fifth order cosmographic developments, supposed in fact to best fit the data,
appear to not well represent the data at high redshifts, say above $z = 4$. The authors themselves disregard two observational values between
$z=7$ and 9, with the argument that the Taylor series are problematic for such high redshifts; however, we do think that 
the problems occur
already at lower values between $z=4$ and 7. 
At these high redshifts, the $\Lambda$CDM model with
$\Omega_{\mathrm{m}}=0.30$ better reproduce the observations than the Taylor series supposed to do it.
The same is true for the SIV with $\Omega_{\mathrm{m}}$
in the range of 0.30 to 0.10.
On the whole, we conclude the scale invariant
models, with $\Omega_{\mathrm{m}}$ between 0.30 and 0.10, are consistent with high redshift GRB data,
although a detailed statistical analysis could perhaps give more precise indications on the value of 
$\Omega_{\mathrm{m}}$ suggested by these observations.


\subsection{The Past Expansion Rates H(z) versus Redshift} \label{HHz}

The expansion rate varies with time in a way depending on the models,
thus the study of $H(z)$ may provide new constraints on cosmological models.
The function $H(z)$ in the SIV theory, for $k=0$, is given by the analytical relation~(\ref{hz2}) seen above.
The first method to study $H(z)$ vs. $z$ was the cosmic chronometer. It is independent of the cosmological models;
however, it depends on assumptions on star formation and evolution. This method has been developed and used 
by a number of authors~\citep{Jim02,Simon05,Stern10,Moresco12,Moresco15,Moresco16,Zhang14,Jesus18}.
The method is based on the simple relation:
\begin{equation}
H(z) =- \frac{1}{1+z} \, \frac{dz}{dt} \, ,
\label{hz3}
\end{equation}
\noindent
obtained from $a_0/a=1+z$ and the definition of $H=\dot{a}/a$. The
ratio $dz/dt$ is estimated from the luminosity and colour evolution of a sample of passive galaxies 
(with ideally no active star~formation) of different redshifts which provides some age estimates 
based on evolution models of stellar populations. 
The assumption on the total absence of star formation (SF) remains a point in discussion,
as well as the initial mass function. It is clear that, if 
there is some undetected SF in a galaxy, its age differences with respect to other galaxies will be biased
and the same for $H(z)$.

Another method is based on the Baryons Acoustic Oscillations (BAO).
In brief, the oscillations result from the propagation of sound waves 
in the early Universe, where photons and baryons are strongly
coupled. These waves have a characteristic scale of about 150 Mpc,
the traces of which remain in the distribution of radiation and matter, observable
in the CMB and in the clustering of matter respectively. The characteristic scale of BAO provides a standard ruler (at the time
of recombination).
After decoupling, the BAO participate in the expansion of the Universe, but their size can be known and thus the observations
of their extension along and transverse to the line of sight 
provides the distance.
The BAO were detected mainly by the two point correlation of quasars and to a smaller extent by the
application of such methods to luminous red galaxies (LRG).
Their distances, redshifts, and $H(z)$ become accessible.
The first detections
were performed in 2005 by the SDSS and 2dFGRS collaborations~\citep{Eisenstein05,Cole05}; they were followed 
by a number of further studies. The BAO at the highest redshifts of around $z=2.3$ have been studied by
\mbox{\citet{FontRibera14,Delubac15,Blomqvist19}}. Among the recent statistical analyses of BAO data from
the cosmic chronometer method, quasar and LRG distributions, we may quote
\citet{Jesus18b} who found $\Omega_{\mathrm{m}}= 0.256 \pm 0.014$
in the context of the flat $\Lambda$CDM and~\mbox{\citet{Ryan18},\citep{Ryan19}} 
who find that the data mildly favor closed
spatial hypersurfaces. A variety of different models have been compared to BAO oscillations data. 
For example, from combined data,~\citet{Sola17}
find some support for the Running Cosmic Vacuum model, which
assumes that the vacuum energy density (thus~$\Lambda$) and $G$
can be a function of time. 

Figure~\ref{Hvs} shows the observed expansion rates $H(z)$ as a function of $z$, compared to the best fit flat $\Lambda$CDM 
as given by~\citet{Jesus18}. Flat SIV models with different $\Omega_{\mathrm{m}}$ 
are also represented by red lines. For our internal consistency, we have chosen the same 
$H_0 = 74$ km s$^{-1}$ Mpc$^{-1}$ as in Figure~\ref{Hovsage}.
However, the low value of $H_0$ could also be chosen and would give a fine 
adjustment for a slightly lower value of $\Omega_{\mathrm{m}}$. There we enter in the well known debate about the tension
in the $H_0$-determinations.
Globally, a scale invariant model with $\Omega_{\mathrm{m}} \simeq 0.20$ (or slightly lower) appears to be in agreement
with the data from observations. The three accurate observations of BAO oscillations near $z=2.3$ well fit
both the $\Lambda$CDM and SIV models.

\begin{figure}[H]
\centering
\includegraphics[width=.75\textwidth]{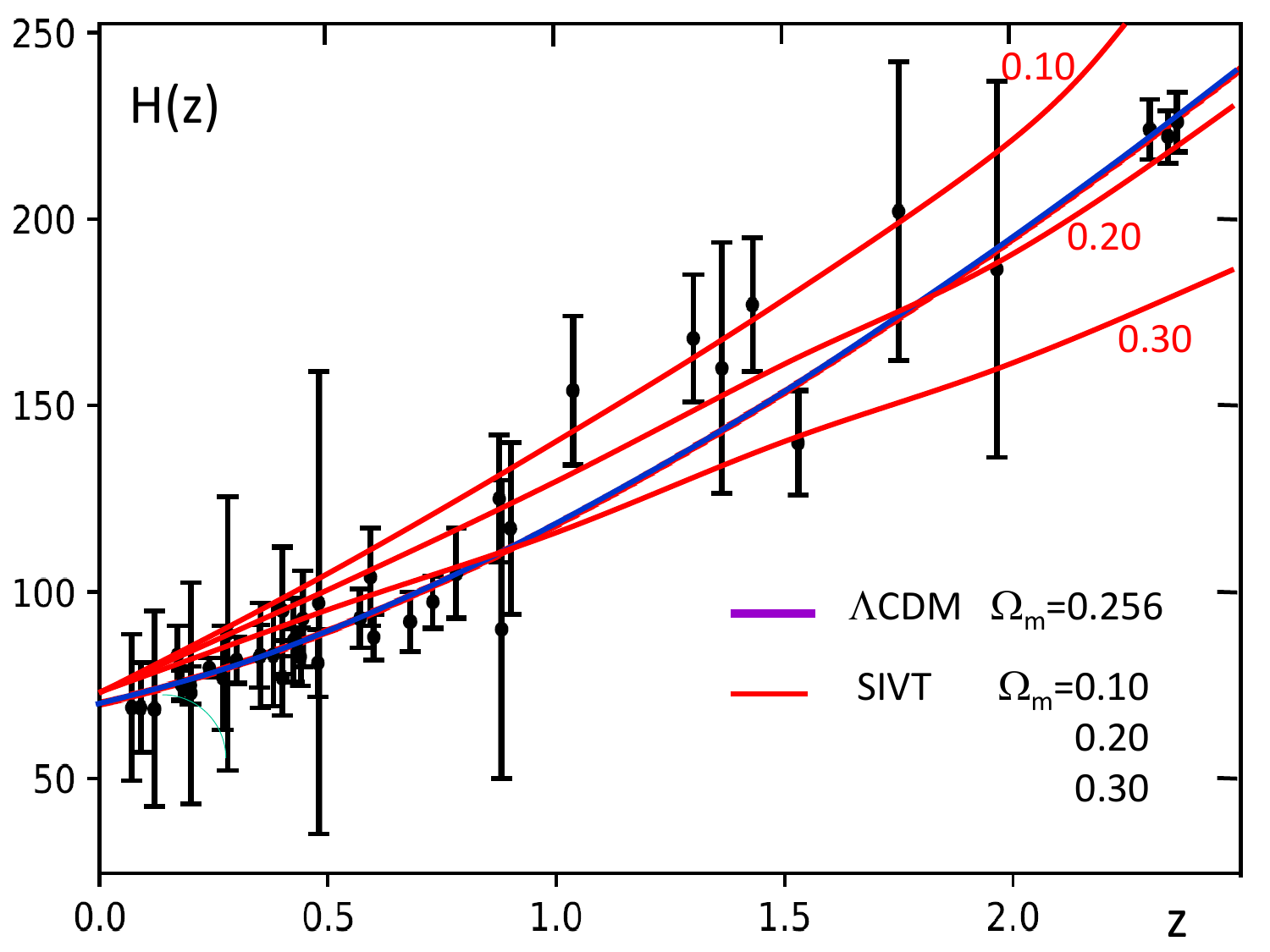}
\caption{The observed expansion rates as a function of redshifts, compared to the flat
$\Lambda$CDM $\Omega_{\mathrm{m}}=0.256$ as given by~\citet{Jesus18b}. The scale invariant models 
with various values of $\Omega_{\mathrm{m}}$ are also represented by red lines.
For consistency within the SIV, and also because it is model independent, we take the value of $H_0$
based on the distance ladder: $H_0 = 74$ km s$^{-1}$ Mpc$^{-1}$ as in Figure~\ref{Hovsage}, while~\citet{Jesus18b} have
$H_0 = 69.5 \;(\pm 2.5)$ km s$^{-1}$ Mpc$^{-1}$.
}
\label{Hvs}
\end{figure}

We may remark that the values of $H(z)$ are not completely independent of 
model assumptions. For example, the value of the standard ruler is fixed by the theory. In the SIV theory, 
the basic equations are different as we have seen it for the growth of density fluctuations~\citep{MaedGueor19}
and the same situation may occur in the study of the oscillations in the initial plasma and their further evolution.
The difference are likely small; nevertheless, this may introduce a possible small bias in this test.

On the whole, the results of the different tests in Figures~\ref{Hovsage},~\ref{Betou}--\ref{Hvs} all point towards a similar value of 
$\Omega_{\mathrm{m}} \simeq 0.20$. The most important point is that these fundamental cosmological tests
do not lead to a rejection of the SIV models, which have the major advantage to require neither 
an unknown source of dark energy, nor of dark matter as shown by the dynamical tests mentioned below.

\subsection{The Redshift Drift in the SIV Theory} \label{drift}

The redshift drift, i.e., the change of redshift $z$ of a given distant object over the years, 
may become a major new cosmological test in the coming decades. The development of a new generation of spectrographs like
ESPRESSO by~\citet{Pepe10} may make such a determination possible.

The time shift of the redshift $z$ of a distant galaxy in the observer restframe is according to~\mbox{\citet{McVittie62}} and~\citet{Liske08},
\begin{equation}
\left(\frac{dz}{dt} \right)_0 \, = \, H_0(1+z) - H(z) \,
\label{mc}
\end{equation}
where $H(z)$ is the expansion factor $\dot{a}/a$ at redshift $z\,= \,(a_0/a)-1$. 
We note that Equation~(\ref{mc}) is also verified in the scale invariant context, since the scale factor 
$\lambda(t)$ applies to both time and space coordinates. 
Thus, the relation $\frac{\Delta t}{\Delta t_0} = \frac{a}{a_0}$ used by McVittie in the demonstration is also valid here.
We have seen in Section~\ref{ddist} that the expansion factor $H(z)$ as a function of redshift can be written, as in Equation~(\ref{hz2})
\begin {equation}
H(z) \, = \, \frac{2 \, \sqrt{1+z}}{1- \Omega_{\mathrm{m}}} \, \left( \Omega_{\mathrm{m}}\,(1+z)^{3/2}+
(1- \Omega_{\mathrm{m}}) \right) ^{2/3} \, .
\end{equation}
\noindent

Thus, the expression of the shift of redshift given by~(\ref{mc}) becomes 
\begin{eqnarray}
\left(\frac{dz}{dt} \right)_0 =H_0(1+z) - H(z) 
= \, H_0\,(1+z) \left[1 - \frac{\left(\Omega_{\mathrm{m}}\,(1+z)^{3/2}+
(1- \Omega_{\mathrm{m}}) \right) ^{2/3}}{(1+z)^{1/2}}\right] \, .
\end{eqnarray}

As such, this equation gives the shift of $z$ in the timescale where the present time is \mbox{$t_0=1$}. To~obtain it for example in years,
one has to express the present Hubble constants in the appropriate~units
\begin{equation}
H_0 \, = \, 70 \, [\frac{km}{s \; Mpc} ]= \, 2.2685 \cdot 10^{-18} \, [ s^{-1}]\,=
\,7.159 \cdot 10^{-11}\, [yr^{-1}] \,.
\end{equation} 
\noindent

We obtain the final expression for the shift
\begin{equation}
\left(\frac{dz}{dt} \right)_0 \, =\,7.159 \cdot 10^{-11}\,h_{70}\,
(1+z) \left[1 - \frac{\left(\Omega_{\mathrm{m}}\,(1+z)^{3/2}+
(1- \Omega_{\mathrm{m}}) \right) ^{2/3}}{(1+z)^{1/2}}\right] \, [yr^{-1}] \,.
\label{final}
\end{equation}

There, $h_{70}$ is the ratio of the chosen value of $H_0$ with respect to 70 km \, s$^{-1}$ \, Mpc$^{-1}$.

Figure~\ref{overall} shows the curves of the shifts as a function of the redshift $z$ predicted in the SIV cosmology
for different values of $\Omega_ {\mathrm{m}}$. They are compared to some standard models of different values of
$\Omega_{\Lambda}$ and $\Omega_ {\mathrm{m}}$ by~\citet{Liske08}.
We notice the relative proximity of the standard and scale invariant curves in the case of $\Omega_ {\mathrm{m}}= 0.30$,
which could make the separation of models difficult. However, 
the expected value of $\Omega_ {\mathrm{m}}$ in the SIV cosmology is likely significantly smaller than in the 
$\Lambda$CDM models, being of the order of 
$\Omega_ {\mathrm{m}}= 0.20$ or even less; this makes the differences of the $z$-drifts between the 
two kinds of cosmological models possibly observable by very accurate observations in the future.

\begin{figure}[H]
\centering
\includegraphics[width=.90\textwidth]{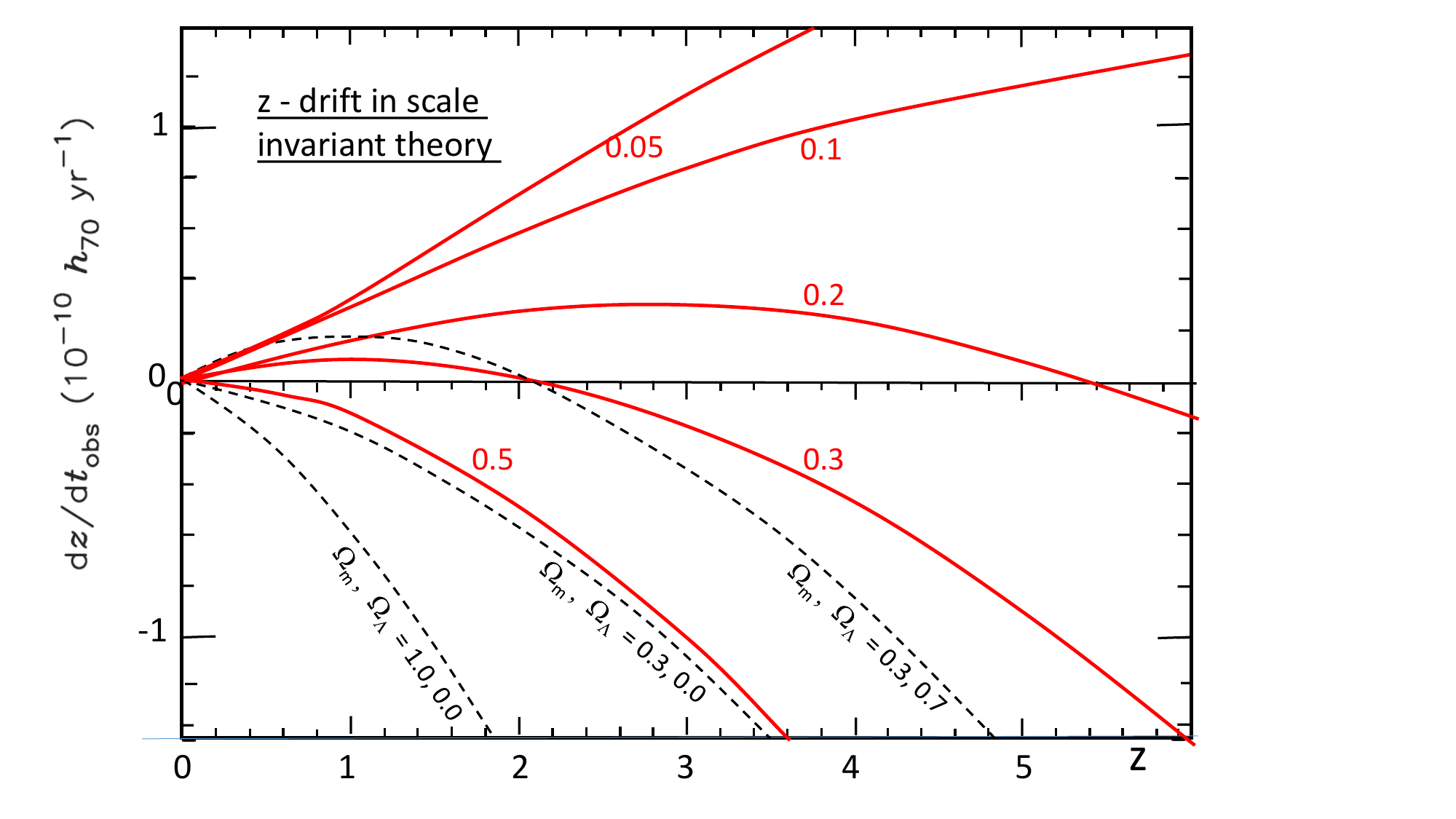}
\caption{The drifts of redshifts $dz/dt$ as a function of redshift in the scale invariant theory (red curves). 
The values of $\Omega_{\mathrm{m}}$ 
(usual definition) are indicated. The black broken lines give the results for some standard models of different couples 
$(\Omega_{\Lambda}, \Omega_{\mathrm{m}})$ by~\citet{Liske08}.}
\label{overall}
\end{figure} 

\section{Properties and Tests in the Newton Like Approximation} \label{weakf}

In this section, we briefly mention the consequences of the SIV hypothesis in the weak field approximation 
and summarize some positive observational tests. Indeed, this shows the similarity of the effects in cosmology
and in the Newtonian approximation. In both cases, the SIV hypothesis results in an additional acceleration in the direction
of the motions. The reason why will be explicated after the basic Equation ~(\ref{Nvec}).
In cosmology, these effects are currently interpreted in terms of dark energy, while, in the Newtonian approximation,
they are accounted for in terms of the dark matter (DM) hypothesis.

\subsection{The Basic Dynamical Equation and the Two-Body Problem} \label{dyn}

The SIV theory leads to important dynamical effects.
The starting point is the geodesic equation, obtained by 
\citet{Dirac73} and also demonstrated by an action principle by 
\citet{BouvierM78}; see also~\citep{MBouvier79, Maeder17c},
\begin{equation}
\frac{du^{\rho}}{ds}+ \Gamma^{\rho}_{\mu \nu} u^{\mu} u^{\nu} -\kappa_{\mu}u^{\mu} u^{\rho}+ \kappa^{\rho} = 0 \, ,
\label{geod}
\end{equation}
\noindent
with the velocity $u^{\mu} \, = \, dx^{\mu}/ds$---in the weak field non-relativistic approximation, with a metric given by
$ g_{i \, i} = -1\, , \; \mathrm{for} \; i=1, 2, 3 \quad \mathrm{and} \quad g_{00}=1+ (2\Phi/c^2) $. For the potential, we
take that of a central mass point $M$ at a distance $r$. 
The Christoffel symbols become: $ \Gamma^i_{00} \,=\, \frac{1}{2} \, \frac{\partial g_{0 0 }}{\partial x^i}\, =
\,\frac{1}{2} \, \frac{\partial \left(1+ (2\Phi/c^2) \right)}{\partial x^i} \, =
\frac{1}{c^2} \frac{\partial \Phi}{\partial x^i} $. With $ds \approx c dt$ and the velocities $u^i \approx \frac{v^i}{c} = \frac{dx^i}{c dt}$ and $u^0 \approx 1$, one obtains
\begin{equation}
\frac{dv^i}{dt}+ \frac{\partial \Phi}{\partial x^i } - \kappa(t) v^i \, = \, 0 \, .
\label{geod22}
\end{equation}

From Equation~(\ref{ro2}), the density is a coscalar of power $\Pi(\rho)=-2$, thus $\frac{M}{r^3} \, = \, \frac{M'}{r'^3} \, \lambda^2$ and
with $r'= \lambda \, r$, we have $M' = \lambda M$. The number of particles forming $M$ is evidently not changing, but 
the inertial properties are changing, a situation also common in Special Relativity. However, 
the gravitational potential is a scale invariant quantity and is thus left unmodified,
$\Phi' \, =- \frac{GM'}{r' } \, = \,- \frac{GM}{r } =\, \Phi $. Equation~(\ref{geod22}) becomes in spherical
coordinates, 
\begin{equation}
\frac {d^2 \vec{r}}{dt^2} \, = \, - \frac{G \, M}{r^2} \, \frac{\vec{r}}{r} + \, \kappa(t)\, \frac{d\vec{r}}{dt} \, ,
\label{Nvec}
\end{equation}
with $\kappa(t)=1/t$. In case of expansion, there is an additional outwards acceleration opposed to the Newtonian gravity;
in case of infall, the additional term adds its effect to the Newtonian attraction. This equation has the same meaning, but
in a different context, as Equation ~(\ref{E3}) where we also see that the total acceleration is the sum of two terms: the Newtonian one
and an additional one in the direction of the velocity. To explain it simply, let us say that 
the additional acceleration is the result of the negative pressure exerted
by the scale invariant empty space. A positive pressure would inhibit the motions, a negative one is favoring them. The scale invariance 
of the equation of motion (geodesics) implies a time dependence of this negative pressure.
The additional term is very small, the inverse of the age 
of 13.8 Gyr is $2.295 \times 10^{-18}$ s$^{-1}$ or 70.85 km s$^{-1}$ Mpc$^{-1}$, close to the current Hubble constant $H_0$.

It is interesting to notice that the ratio $x$ of the additional acceleration to Newtonian gravity behaves to first order 
like~\citep{Maeder17c,MaedGueor20}, 
\begin{equation}
x \, \geq \, \frac{\sqrt{2}}{\xi}\,\left( {\frac{\varrho_{\mathrm{c}}}{\varrho }} \right)^{1/2} \, ,
\label{x}
\end{equation}
where $\xi=H(t)/\kappa(t)$ is of order $1$ and $\varrho_{\mathrm{c}}$ is the critical density in~(\ref{defomega}). 
For relatively dense systems with $\varrho > 10 \varrho_{\mathrm{c}}$, the equality in the above expression is closely satisfied.

The two-body problem has been studied by~\citet{MBouvier79} and~\citet{Maeder17c}.
The~equation of motion~(\ref{Nvec}) writes in the two polar coordinates
\begin{equation}
\ddot{r} - r \, \dot{\vartheta}^2 \, = \,- \frac{G \, M}{r^2} +\kappa(t) \, \dot{r} \, ,
\quad \quad \mathrm{and} \quad
r \, \ddot{\vartheta} + 2 \, \dot{r} \, \dot{\vartheta} \, = \, \kappa(t) \, r \, \dot{\vartheta} \, .
\label{Ntheta}
\end{equation}

The second of these equations implies the following law of angular momentum conservation
\begin{equation}
r^2 \, \dot {\vartheta} \frac{t_0}{t} \, = \, L \, = \, \mathrm{const.}
\label{ang}
\end{equation}

This equation with the first one in~(\ref{Ntheta}) leads to a modified form of Binet equation, the solution of which are conics,
with a constant eccentricity and a secular variation of the semi-major axis \mbox{(see above references)}. From this last relation, 
we see that the velocity of circular motions behaves like,
\begin{equation}
v_{\mathrm{circ}} \, = \, r \, \dot{\vartheta} \, = \, \frac{ L }{r} \, \frac{t}{t_0} \, .
\label{vcirc}
\end{equation}

Since the lengths increase with time, this expression implies that the circular velocity in the case of the two-body motion 
is a scale invariant, a fact which is consistent with the inscalar property of the potential found in
Section~\ref{basicprop}. Indeed, the circular velocity being an inscalar will write
\begin{equation}
v^2_{\mathrm{circ}} \, = \, \frac{G \, M'}{r'} \, = \, \frac{G \, M}{r} \,.
\label{vcirc}
\end{equation}

Thus, the circular radius will behave like
\begin{equation}
r \, = \, \frac{GM}{v^2} \, = \, \frac{GM \, r^2}{L^2} \left(\frac{t_0}{t}\right)^2 \quad \mathrm{giving} \quad
r \, = \, \frac{L^2 t^2}{G \, M t^2_0} \, = \, \frac{L^2 t}{G \, M' t_0} \, .
\end{equation}

The orbital radius of circular motions, and the same for the semi-major axis $A$, not to be confused with the scale factor $a$, 
of the conics, experiences a secular increase so that the motions are in reality slow outwards spiraling motions. 
Due to the conservation law~(\ref{ang}) of the
angular momentum, the rotation period $T$
is also changing the same way~\citep{MBouvier79},
\begin{equation}
\frac{\dot{A}}{A} \, = \, \frac{1}{t } \, , \quad \mathrm{with \;~(\ref{ang}) \; 
it \; gives} \quad \frac{\dot{T}}{T} \, = \, \frac{1}{t } \, .
\label{aT}
\end{equation}

If we consider a test mass around a central object of mass $M$,
the additional term in~(\ref{Nvec}) produces an acceleration proportional and along to the rotation velocity
with no component along the Newtonian gravity. 
This acceleration leads to the above expansion of the orbit. At the same time, the above acceleration together
with some slowing down due to the angular momentum properties leads to a conservation of the orbital velocity.
Indeed, we verify in Equations~(\ref{aT}) that the semi-major axis and the rotation period are varying
in the the same way, consistently with a constant orbital velocity and also with the expression of the
gravitational potential (Section~\ref{basicprop}).

\subsection{Tests on Stellar Dynamics and Galaxies}

We briefly summarize here the main points about the tests already performed, as applications 
of the modified Newton law~(\ref{Nvec}). They concern 
clusters of galaxies, galaxy rotation, and the ``vertical'' velocity dispersion of stars in the Milky Way.

\subsubsection{Clusters of Galaxies}
Clusters of galaxies show mass to light ratios $M/L$ which are currently 
much larger than the value expected for a typical stellar population
($\sim$ 10 M$_{\odot}$/L$_{\odot}$). 
As an example, for 600 clusters in the range of $10^{14}$ to $10^{15}$ M$_{\odot}$, $M/L$ ratios in the range of
300--500 M$_{\odot}$/L$_{\odot}$ have been found~\citep{Proctor15}. 
The problem within the SIV context has been discussed
by~\citet{Maeder17c}, where further references and details are given. 
Writing~(\ref{Nvec}) for a given galaxy in a cluster and multiplying by 
$\upsilon_i$ leads to
\begin{equation}
\frac{1}{2} \, d (\upsilon^2_i ) \, = 
- \sum_{j \neq i} \frac{G \, m_j dr_{ij}}{r^2_{ij}} + \kappa(t) \, \upsilon^2_i \, dt \, ,
\label{imd}
\end{equation}

Integrating and taking the mean leads to
\begin{equation}
\overline{\upsilon^2} \approx q' \frac{GM}{R} + 2 \int \kappa(t) \overline{\upsilon^2} dt \,
\label{v2}
\end{equation}
with $q'$ a numerical factor. The ratio of the last term to $\overline{\upsilon^2}$ is of the order of 
$2 (R/t) \, \overline{\mid \upsilon \mid}$. 
If the clusters are not too distant, $t$ is of the order of the age of the Universe. We see that with Equation ~(\ref{v2}) a smaller mass is
associated with a given velocity dispersion $\overline{v^2}$ than in the standard theory where the last term is absent.
The effects reach an order of a magnitude and more depending on the mean cluster density.

\subsubsection{Galaxy Rotation and the RAR} \label{RAR}

Spiral galaxies rotate faster in their outer layers than would be expected from Newton's theory
for the mass present in them~\citep{Sofue01}. This well-known problem, 
at the origin of the dark matter paradigm, is beautifully illustrated by the radial acceleration relation (RAR), 
which compares the dynamical gravity acceleration $g_{\mathrm{obs}} = \upsilon^2/R$
with the matter gravity $g_{\mathrm{bar}}= GM/R^2$ in different points $R$ of late, early and dwarf spheroidal galaxies,
see in particular~\citet{McGaugh16} and~\citet{Lelli17}. There, $\upsilon$ is the circular velocity at a distance $R$ from the center,
$M$ is the mass inside radius $R$. In the SIV theory, a relation between $g_{\mathrm{obs}}$ and $g_{\mathrm{bar}}$ in the case of 
circular motion has recently been obtained by~\citet{MaedGueor20} in the case of galactic circular motions,
\begin{equation}
g_{\mathrm{obs}}\,=\,g_{\mathrm{bar}}+\frac{k^{2}}{2} + \frac{1}{2}\sqrt{4g_{\mathrm{bar}}k^{2}+k^{4}}\,.
\label{sol}
\end{equation}

This equation predicts two limiting cases: 
the Newtonian limit and a limit when $g_{\mathrm{bar}}$ tends towards zero:
\begin{equation}
(1) \;\mathrm{If} \; k^{2} \, \ll \, g_{\mathrm{bar}} \; \mathrm{then}\;g_{\mathrm{obs}} \,\rightarrow\,g_{\mathrm{bar}}\,.
\quad \quad 
(2) \; \mathrm{If}\; g_{\mathrm{bar}}\,\rightarrow \, 0 \, ,\; \mathrm{then}\;g_{\mathrm{obs}}\,\rightarrow\,k^{2}\,.
\label{asy}
\end{equation}

The parameter $k^2$ (potentially a constant ) appears as a background limit of the dynamical gravity for a vanishing Newtonian gravity,
independently of the radius considered.
The background acceleration, $k^2$ is logically related to some cosmological properties of the Universe,
in particular to the average matter density~\citep{MaedGueor20}, 
\begin{equation}
k^{2}\,\approx
\,\frac{1}{2}\,f \, \Omega_{\mathrm{m}}\,c\,H_{0}\,.
\label{k2fin}
\end{equation}

Expression~(\ref{sol}) may also be simplified for gravities $g_{\mathrm{bar}}$ sufficiently larger than $ k^2$. 
One may write $g_{\mathrm{obs}}\,\rightarrow\,g_{\mathrm{bar}}\,+\,\sqrt{g_{\mathrm{bar}}\,k^{2}}$.
This may be compared to MOND, which also has
two limits~\citep{Milgrom83,Milgrom09}, the Newtonian one, and $ g \, \rightarrow \sqrt{g_{\mathrm{N}}\,a_{0}}$, for
$g_{\mathrm{N}}\ll a_{0}$.
There $g_{\mathrm{N}}$ corresponds to $g_{\mathrm{bar}}$ and $g$ to $g_{\mathrm{obs}}$, $a_0 = 1.2 \cdot 10^{-10}$
m s$^{-2}$, which is the so-called deep MOND limit.

Figure~\ref{gobs} shows the comparison of MOND and SIV results. Over the range from the large elliptical
galaxies to the low mass spirals down to the above-mentioned limit, the agreement
between the two curves, as well as with the observations is excellent.
The curves are different for very low gravities, with larger values of $g_{\mathrm{obs}}$ for the SIV results, 
which show agreement with the dwarf spheroidals for the indicated value of $k^2$.
The MOND predictions and the observations
of the dwarf spheroidals do not fit, a point also mentioned by~\citet{Dutton19}.
Thus, the dwarf spheroidals, which are the objects where the amount of dark matter with respect
to baryons is the highest, appear to have a critical role in constraining theories.

\begin{figure}[H]
\centering 
\includegraphics[width=12.0cm, height=8.0cm]{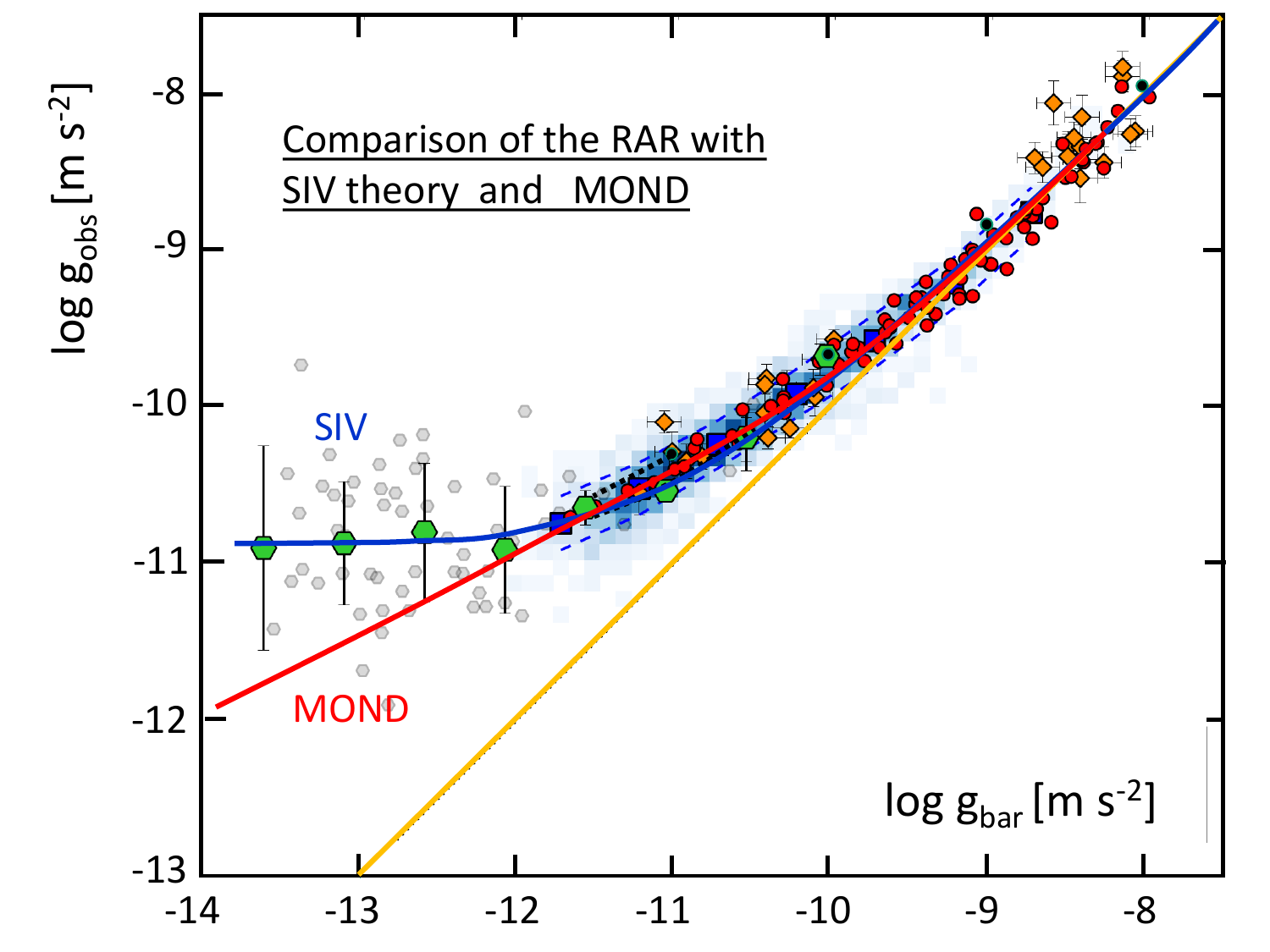}
\caption{Observed values of $g_{\mathrm{obs}}$ and $g_{\mathrm{bar}}$ for
the 240 galaxies studied by~\citet{Lelli17}, forming the radial acceleration
relation (RAR). The big green hexagons represent the binned data of
the dwarf spheroidal galaxies. The blue curve shows the relation predicted by 
(\ref{sol}) with a value of \mbox{$k^{2}=10^{-10.85}$
m s$^{-2}$.} The red curve gives the MOND relation~\citep{Milgrom16}
\mbox{$g\,=\,g_{\mathrm{N}} [1-e^{-(g_{\mathrm{N}}/a_{0})^{\frac{1}{2}}}]^{-1}$}
with \mbox{$a_{0}=1.20\cdot10^{-10}$}
m s$^{-2}$, while the orange curve shows the 1:1-line.}
\label{gobs} 
\end{figure}

The above results have been disputed by~\mbox{\citet{Banik20}} on the basis of the study by~\mbox{\citet{Hees14}} on the motion of the Cassini 
spacecraft around Jupiter. The remark by \mbox{\citet{Banik20}} is poorly founded and lacks a valid scientific basis. Indeed, 
the data reduction by Hees et al. does not apply to the SIV theory. 
The orbit of Saturn and the external field effects (EFE) have been calculated with the MOND equations
and not with the appropriate SIV equation of motion~(\ref{Nvec}).  
Specifically, for the determination of the orbital parameters of Saturn, the SIV theory should be used in the data reduction; 
otherwise, the test is inconsistent. In addition, Equation~(\ref{sol}) for galaxies cannot be applied as such to the Solar System
with the same constant $k^2$. This term represents a background acceleration, 
which is evidently not the same for galaxies and the Solar system.

{ {
We have mentioned in Section~\ref{weyl} the introduction of a vector field by Meierovich~\cite{Meierovich12} ,
which can represent the  properties of dark matter well. This approach has similarities with ours and 
also results in a satisfactory representation of the rotation curves of spiral galaxies~\cite{Meierovich13}, 
and thus finding some justification for the MOND theory. According to Meierovich~\cite{Meierovich19}, the rotation curves also show 
some oscillations with a dependence of the plateau velocity on the mass of the central black hole. 
Such dependence is yet to be investigated within the SIV model and/or to be confirmed observationally.
}}

\subsubsection{The Vertical Dispersion of Stellar Velocities in the Galaxy}

The velocity dispersion of stars in the Milky Way, in particular for the $W$--component (perpendicular to the Galactic plane),
considerably increases with the age of the stars considered; see,~for example,~\citet{Seabroke07}. 
Continuous processes, such as spiral waves, collisions with giant molecular clouds, etc... are active in
the disk plane and may effectively influence the stellar velocity distributions. However, the stars
spend most of their lifetime out of the galactic plane and, in order to have some significant effect,
the stars should receive some ``heating'' also away from the Galactic plane, as shown by these authors. This problem has a long 
history~\citep{Spitzer51} and no consensus has been found on the origin of the effect~\citep{Kumamoto17}.

A study of the oscillations of stars around the Galactic plane in the SIV context shows that an equation like~(\ref{Nvec}) amplifies 
the oscillations. The velocity $W(t_{\mathrm{in}})$ of a star born at time $t_{\mathrm{in}}$,
when crossing the plane, is~\citep{Magnenat78, Maeder17c}, 
\begin{equation}
W(t_{\mathrm{in}}) \, = \, W(t_0) \, \frac{ t_0}{t_{\mathrm{in}}} \, .
\label{W}
\end{equation}
\noindent
where $t_0$ is the present time. The trend for the velocity dispersions follows that of the velocities.
We take a value of 10 km s$^{-1}$ for the present velocity dispersion $\sigma_{\mathrm{W}}$ (cf. Figure~\ref{dispersion}).
As an example for a group of stars with a mean age of 10 Gyr, for an age of the universe of 13.8 Gyr, the velocity dispersion is 
estimated to be about 10 km s$^{-1} \,\times \frac{13.8}{3.8} = 36.3$ km s$^{-1}$.
Figure~\ref{dispersion} compares the corresponding model predictions obtained in this way (continuous red curve) with the data from~\citet{Seabroke07}. 
We see that the theoretical curve 
well corresponds to the trend shown by the observations. 

\begin{figure}[H]
\centering
\includegraphics[width=.50\textwidth]{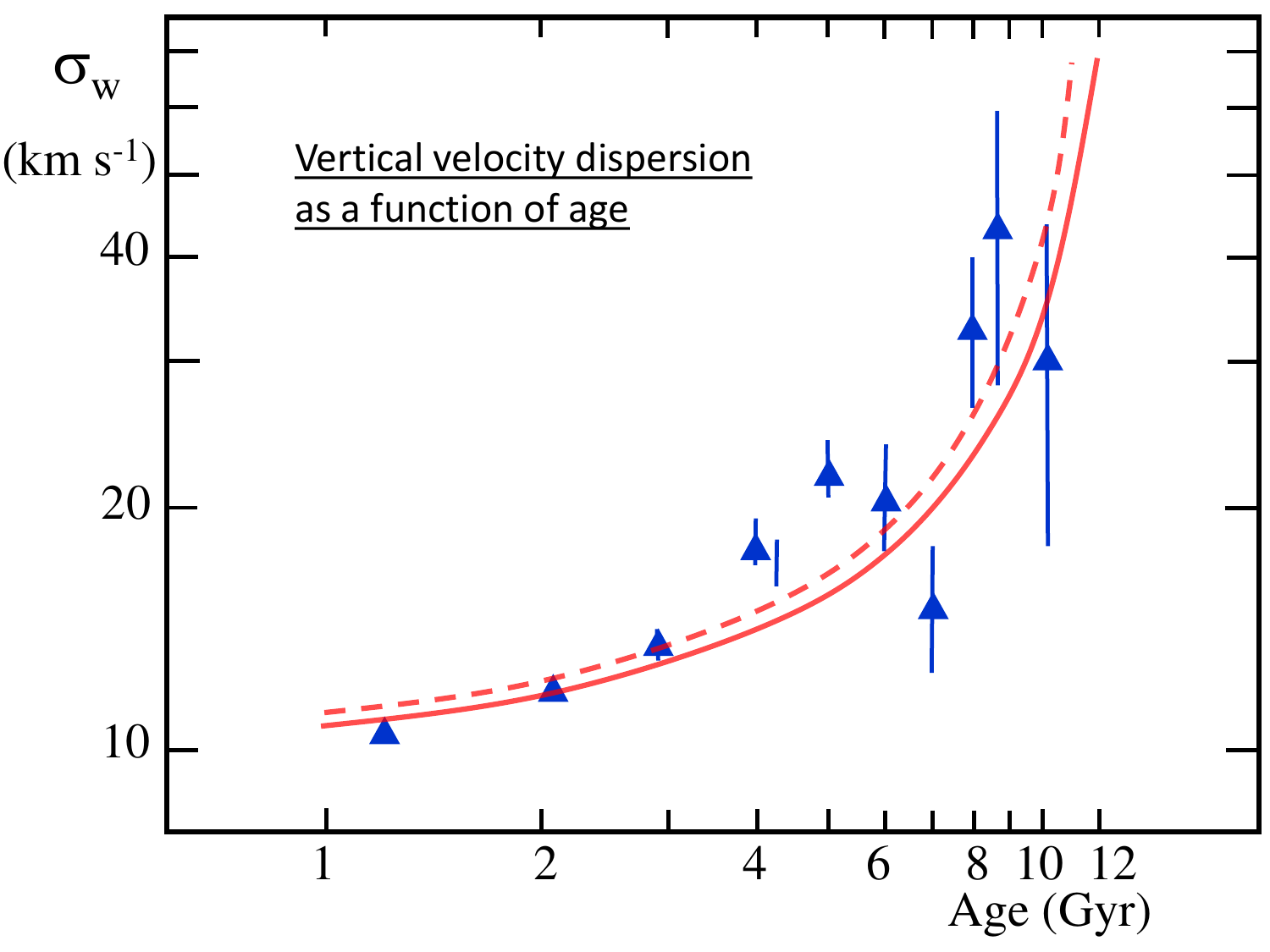}
\caption{The vertical dispersion $\sigma _W$ as a function of the age of the stellar populations. The blue triangles with 
the error bars result from the analysis by~\citet{Seabroke07} of the observations by~\citet{Nordstrom04}. 
The continuous red curve shows the predictions of the scale invariant dynamics according to relation~(\ref{W}) for an age 
of the universe of 13.8 Gyr. The broken red curve accounts for the fact that the Galaxy formed about 400 Myr after the 
Big Bang. A vertical dispersion of 10 km s$^{-1}$ is assumed at the present time. }
\label{dispersion}
\end{figure}

\subsection{The Growth of the Density Fluctuations} \label{drho}

The growth of density fluctuations in the early Universe is a fundamental cosmological problem and
one of the main reasons, along with the flat rotation curves of spiral galaxies, for the introduction 
of dark matter DM~\citep{Coles02,Durrer08}. 
At recombination ($z \sim 10^3$),
the observed fluctuations of the CMB are of the order of $\delta T/T \approx 10^{-5}$.
At this time, the fluctuations of the baryonic density $\delta \rho/\rho$
are still of the same order, since, up to recombination, matter and radiation are closely coupled. 
From this time onward, the growth of the density fluctuations $\delta \, \equiv \, \delta \rho/\rho$ 
is determined by gravitation and expansion~\citep{Peebles80}.
In~the standard model,
$\delta$ goes like $t^{2/3} \, \propto \, a(t)\, \propto \, (1+z)^{-1}$. Thus, starting with an amplitude of
about~$10^{-5}$, a growth by a factor of $10^3$
would lead to fluctuations of about $10^{-2}$ at the present epoch. 
This is by several orders of magnitude smaller than the large nonlinear structures presently observed in the Universe 
\citep{Ostriker93}.

According to the standard model, during the radiative era,
the DM perturbations (not subject to radiation pressure) would be growing, while the baryonic perturbations do not amplify.
After recombination,
the baryons fall into the potential wells of the previously assembled supposed
dark matter, thus giving an extraordinary boost to the growth of the baryonic density fluctuations. 
This~scenario of the cold dark matter (CDM) is a basis of the
standard $\Lambda$CDM model.

Classically, the equations of Euler and Poisson, as well as the continuity equation, 
form the basis for the study of the evolution of the density fluctuations.
The scale invariant form of the Euler equation becomes~\citep{MaedGueor19}
\begin{equation}
\frac{d\vec{v}}{dt}=\frac{\partial\vec{v}}{\partial t}+\left(\vec{v}\cdot\vec{\nabla}\right)\vec{v}=
-\vec{\nabla}\Phi-\frac{1}{\rho}\vec{\nabla}p+\kappa\vec{v}\label{Euler} \, .
\end{equation}

The density $\varrho$ is the baryon density.
It contains an additional term proportional to the velocity, as in Equation~(\ref{Nvec}).
This term plays an important role in the enhancement of density fluctuations in the SIV theory. 
The corresponding continuity equation is:
\begin{equation}
\frac{\partial \rho}{\partial t}+ \vec{\nabla}\cdot (\rho \vec{v}) = \kappa \left [\rho+ \vec{r} \cdot \vec{\nabla} \rho \right]\, .
\end{equation}

The second member can be viewed as a diffusion term. The potential being scale invariant, 
the Poisson equation $\vec{\nabla}{}^{2}\Phi=\triangle\Phi=4\pi G\varrho $ stays the same.
These three equations applied to a density perturbation $\delta = \delta \varrho/\rho$ in the linear regime
lead to the basic equation for the growth of density perturbations in the SIV theory~\citep{MaedGueor19}, 
\begin{equation}
\ddot{\delta}+ (2H -(1+n) \kappa)\dot{\delta} \, =
\, 4\pi G \varrho \delta + 2 n \kappa (H-\kappa) \delta \, .
\label{D2}
\end{equation}

Here, $H$ is the Hubble expansion rate at the time considered, $n$ is related to the slope of the density profile of the perturbation,
$n= d \ln \delta/d \ln x$, where $x$ is the distance {\it{to}} the perturbation . The NFW profile, which in the outer layers
of a cluster is very close to the $\varrho \sim 1/r^2$ profile~\citep{NFW95} would have a value $n=2$. 
The terms with $\kappa= - \dot{\lambda}/\lambda$ are absent in the standard models (there $\lambda=1$) where one 
has the well-known equation,
\begin{equation}
\ddot{\delta}+ 2H \dot{\delta} \, = \, 4\pi G \varrho \delta \, .
\label{D3}
\end{equation}

Figure~\ref{contn2} shows some results. The main result is that the growth of the density fluctuations $\delta$
is fast enough in the SIV cosmology to account for the formation of galaxies long enough before the present. There is no need 
for the dark matter hypothesis to favor structure formation.
Another point we notice is that the growth of $\delta$ is faster for lower $\Omega_{\mathrm{m}}$ values. This result, 
surprising at first glance, is consistent with the fact 
that the scale invariant effects are larger at lower densities as shown by Equation ~(\ref{x}).

\begin{figure}[H]
\centering
\includegraphics[width=.65\textwidth]{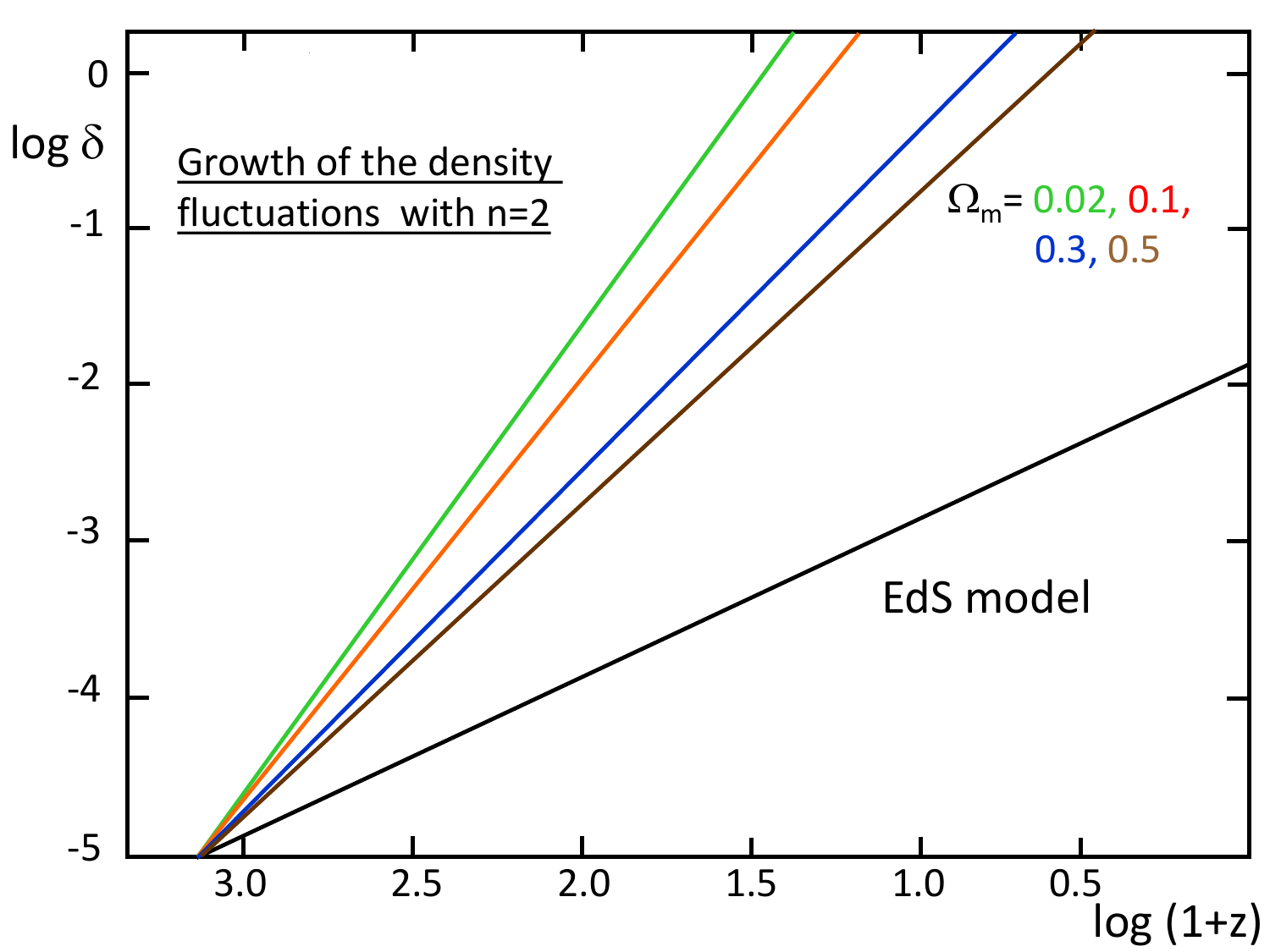}
\caption{The growth of density fluctuations for different values of parameter $\Omega_{\mathrm{m}}$ for $n=2$.
The~values $\delta=1$ are reached in the interval of redshifts $(z+1)=4.0$ to 29.3. From~\citep{MaedGueor19}.}
\label{contn2} 
\end{figure} 

The results of Figure~\ref{contn2} are obtained for a value $n=2$. Higher values such as $n=3$ or 5 produce
a faster growth of the density fluctuations~\citep{MaedGueor19}. 
For example, for $\Omega_{\mathrm{m}}= 0.10$ and $n=2$, $\delta=1$ is reached at $z=10$,
at $z \approx 25$ for $n=3$, and at $z \approx 80$ for $n=5$, which is certainly too high.
A recent observation of a gravitationally lensed galaxy shows 
signatures of star formation (OII lines) implying an active star formation
at about $z=15$~\citep{Hashimoto18}. Thus, also depending on the exact value
of $\Omega_{\mathrm{m}}$, a slighltly steeper concentration than that of the typical $n=2$ model
may lead to an early star formation in the Universe.
This is also a positive feature of the SIV models.

\section{Conclusions and Perspectives} 
\label{concl}
The SIV hypothesis that the macroscopic empty space is scale invariant
leads to an additional acceleration of motions, 
which in cosmology produces effects currently attributed to the dark energy,
while, in the Newtonian-like mechanics, it gives effects attributed to the dark matter.
At this stage, there is already a number of favorable tests of the SIV theory both in cosmology and
in the study of astrophysical motions. These results do not constitute a proof of the theory. 
However, they are sufficiently promising to justify further investigations of the SIV theory and
its properties.

\vspace{6pt}
\authorcontributions{
A.M, is the lead author and researcher on the topics discussed by the paper.
He has analyzed the relevant data and has produced and published previously many of the graphs shown in the Figures.
V.G. has verified the mathematical correctness of the equations and the relevant physics conclusions. 
He has independently validated the trends depicted in the graphs that have been part of the previously coauthored papers with A.M.
Both co-authors have been actively involved in the writing of the paper and its draft versions (overall more than a dozen). }  


\acknowledgments{A.M. expresses his deep gratitude to his wife and to D. Gachet for their continuous support.
V.G. is extremely grateful to his wife and daughters for their understanding and family support 
during the various stages of the research presented. This research did not receive any specific 
grant from funding agencies in the public, commercial, or not-for-profit sectors.}

\funding{This research received no external funding.}

\conflictsofinterest{The authors declare no conflict of interest.} 


\appendixtitles{yes} 
\appendix

\section{The Integrable Weyl Geometry} \label{hweyl}

In addition to the general covariance of General Relativity, 
Weyl's Geometry also considers gauge or scale transformations.
It was first developed by Hermann Weyl~\citep{Weyl23} and further studied by
\citet{Eddington23} and~\citet{Dirac73}.
The Integrable Weyl's Geometry is a particularly interesting case of this geometry 
as shown by~\citet{Canu77} and~\citet{BouvierM78}, as discussed below.
The original aim of Weyl was to interpret the electromagnetism in
term of properties of the space-time geometry, as Einstein did for
gravitation. This geometry is endowed with a metric determination
of the quadratic form $ds^{2}\,=\,g_{\mu\nu}(x)dx^{\mu}dx^{\nu}$
\, as in Riemann space, and with quantities expressing gauge transformations.
Let us consider a vector of length $\ell$ attached at a point $P$
of coordinates $x^{\mu}$. If this vector is transported by parallel
displacement to a point $P'$ of coordinates $x^{\mu}+\delta x^{\mu}$,
its length becomes $\ell+\delta\ell$, with
\begin{equation}
\delta\ell\,=\,\ell\kappa_{\mu}\delta x^{\mu}\,
\label{kk1}
\end{equation}
where $\kappa_{\mu}$ is the coefficient of metrical connexion.
In Weyl's geometry, the terms $\kappa_{\mu}$ are fundamental coefficients
as are the $g_{\mu\nu}$ in GR. Weyl considers that the lengths
can undergo gauge changes: 
\begin{equation}
\ell'\,=\,\lambda(x)\,\ell\,\label{l1}
\end{equation}
where $\lambda$ is the scale or gauge factor, which could in principle depend on the  four coordinates. 
The~segment $\delta\ell'$
undergoes gauge transformation changes as well. To the first order
in $\delta x^{\nu}$, one has: 
\begin{equation}
\ell'+\delta\ell'\, = \,(\ell+\delta\ell)\lambda(x+\delta x)\approx
(\ell+\delta\ell)\lambda(x)+\ell\frac{\partial\lambda}{\partial x^{\nu}}\delta x^{\nu}\,
\end{equation}
\begin{eqnarray}
\delta\ell'\, &=& \,\lambda\delta\ell+\ell\lambda_{,\nu}\delta x^{\nu}=
\lambda\ell\kappa_{\nu}\delta x^{\nu}+\ell\lambda_{,\nu}\delta x^{\nu} \,= \\ \nonumber
&=& \lambda\ell\left(\kappa_{\nu}+\Phi_{,\nu}\right)\delta x^{\nu}\,=l'\kappa'_{\nu}\delta x^{\nu}\,
\end{eqnarray}
where the notation $\lambda_{,\nu}=\frac{\partial\lambda}{\partial x^{\nu}}$
is used along with: 
\begin{equation}
\Phi\,=\ \ln\lambda\,\quad\quad\mathrm{and}
\quad\kappa'_{\nu}=\kappa_{\nu}+\Phi_{,\nu}=\kappa_{\nu}+\partial_{\nu}\ln\lambda\,.
\label{Phi}
\end{equation}

If the vector is parallel transported along a closed loop, the total
change of the length of the vector can be written as: 
\begin{equation}
\Delta\ell\,=\,\ell\left(\partial_{\nu}\kappa_{\mu}-\partial_{\mu}\kappa_{\nu}\right)\,\sigma^{\mu\nu}\,\label{boucle}
\end{equation}
where $\sigma^{\mu\nu}=dx^{\mu}\wedge dx^{\nu}$ is an infinitesimal
surface element corresponding to the edges $dx^{\mu}$ and $dx^{\nu}$.
The tensor $F_{\mu\nu}\,=\,\left(\kappa_{\mu,\nu}-\kappa_{\nu,\mu}\right)$
was identified by Weyl with the electromagnetic field. However, in
the above form of Weyl's geometry, the lengths are non-integrable:
the change of the length of a vector between two points depends on
the path considered. Such a property would imply that different atoms,
due to their different world lines, would have different properties
and thus will emit at different frequencies. This was the essence
of Einstein's objection against Weyl's geometry, as recalled by \mbox{\citet{Canu77}}.

The above objection does not hold if one considers the 
Integrable Weyl's Geometry, which forms a consistent framework for
the study of gravitation as emphasized by~\mbox{\citet{Canu77}} and~\mbox{\citet{BouvierM78}}.
Let us consider that the framework of functions denoted by primes
is the Riemann space as in GR. The line element
$ds'=g'_{\mu\nu}dx^{\mu}dx^{\nu}$ refers to GR, while $ds$ as in 
Equation~(\ref{ds}) refers to the Integrable Weyl Geometry, which in addition
to the quadratic form of the $g_{\mu\nu}$ also has a
scalar gauge field $\lambda$.
In this case, the interesting point is that we have $\kappa'_{\nu}=0$, and therefore:
\begin{equation}
\kappa_{\nu}\,=\,-\Phi_{,\nu}\,=\,-\frac{\partial\ln\lambda}{\partial x^{\nu}}\,.
\label{k3}
\end{equation}

The above condition means that the metrical connexion $\kappa_{\nu}$
is the gradient of a scalar field \mbox{$(\Phi=\ln\lambda)$}, as seen above.
Thus, $\kappa_{\nu}dx^{\nu}$ is an exact differential and therefore:
\begin{equation}
\partial_{\nu}\kappa_{\mu}\,=\,\partial_{\mu}\kappa_{\nu},\,\label{dx}
\end{equation}
which, according to Equation~(\ref{boucle}), implies that the parallel
displacement of a vector along a closed loop does not change its length;
equivalently, the change of the length due to a displacement does not
depend on the path followed. The mathematical tools of Weyl's geometry
also work in the integrable form of this geometry and this is what we are
using in this work.




\reftitle{{References}} 




\end{document}